\documentclass[aps,prd,final,nofootinbib,twocolumn
]{revtex4}

\usepackage{hyperref}
\usepackage{graphicx}                            
\usepackage{amsmath} 
\usepackage{perpage} 
\usepackage[T1]{fontenc} 
\usepackage{amssymb}
\usepackage{graphicx}                            
\usepackage{amsmath} 
\usepackage{perpage} 
\usepackage{color}
\usepackage{cleveref}
\usepackage{feynmf}
\usepackage{upgreek}
\usepackage{listings}
\usepackage{ctable}
\usepackage{lipsum}

\definecolor{dkgreen}{rgb}{0,0.6,0}
\definecolor{gray}{rgb}{0.5,0.5,0.5}
\definecolor{mauve}{rgb}{0.58,0,0.82}

\newcommand{\avg}[1]{\left\langle#1\right\rangle}

\newcommand{\abs}[1]{\lvert#1\rvert}

\newcommand*\colvec[3][]{\begin{pmatrix}\ifx\relax#1\relax\else#1\\\fi#2\\#3\end{pmatrix}}
\newcommand{\beq}{\begin{equation}}
\newcommand{\beqn}{\begin{eqnarray}}
\newcommand{\eeq}{\end{equation}}
\newcommand{\eeqn}{\end{eqnarray}}
\newcommand{\pd}[2]{\frac{\partial #1}{\partial #2}} 
\newcommand\numberthis{\addtocounter{equation}{1}\tag{\theequation}}
\newcommand{\dbar}{\ensuremath{\mathchar'26\mkern-12mu d}}

\renewcommand{\vec}[1]{\mathbf{#1}}
\newcommand{\bq}{\mathbf{q}}
\newcommand{\nn}{\nonumber}
\newcommand\order[1]{{\cal O}#1}

\DeclareRobustCommand{\Eq}[1]{Eq.~(\ref{#1})}
\DeclareRobustCommand{\Eqs}[2]{Eqs.~(\ref{#1}) and (\ref{#2})}
\DeclareRobustCommand{\Eqss}[3]{Eqs.~(\ref{#1}), (\ref{#2}), and (\ref{#3})}
\DeclareRobustCommand{\Sec}[1]{Sec.~\ref{#1}}
\DeclareRobustCommand{\App}[1]{Appendix~\ref{#1}}
\DeclareRobustCommand{\Fig}[1]{Fig.~\ref{#1}}

\begin{document}
\title{Non-Gaussian Covariance of the Matter Power Spectrum in the Effective Field Theory of Large Scale Structure}
\author{Daniele Bertolini}
\author{Katelin Schutz}
\author{Mikhail P. Solon}
\author{Jonathan R. Walsh}
\author{Kathryn M. Zurek}
\affiliation{Berkeley Center for Theoretical Physics, University of California, Berkeley, CA 94720 \\
Theoretical Physics Group, Lawrence Berkeley National Laboratory, Berkeley, CA 94720}
\begin{abstract}
\noindent
We compute the non-Gaussian contribution to the covariance of the matter power spectrum at one-loop order in Standard Perturbation Theory (SPT), and using the framework of the effective field theory (EFT) of large scale structure (LSS). The complete one-loop contributions are evaluated for the first time, including the leading EFT corrections that involve seven independent operators, of which four appear in the power spectrum and bispectrum. We compare the non-Gaussian part of the one-loop covariance computed with both SPT and EFT of LSS to two separate simulations.  In one simulation, we find that the one-loop prediction from SPT reproduces the simulation well to $k_i + k_j \sim$ 0.25 h/Mpc, while in the other simulation we find a substantial improvement of EFT of LSS (with one free parameter) over SPT, more than doubling the range of $k$ where the theory accurately reproduces the simulation. The disagreement between these two simulations points to unaccounted for systematics, highlighting the need for improved numerical and analytic understanding of the covariance.

\end{abstract}

\maketitle
\section{Introduction}
In the era of precision cosmology, understanding the formation of LSS is essential for gaining insight into physics beyond the Standard Model and of the primordial universe. To that end, a wide range of ongoing and upcoming surveys are leveraging the synergy between different probes of LSS to constrain properties of, for instance, inflation, dark energy, and massive neutrinos \cite{2012SPIE.8446E..0ZM,2005astro.ph.10346T,2013Msngr.154...44J,2012arXiv1211.0310L,2013AJ....145...10D,2013arXiv1308.0847L,2015arXiv150804473D,2011arXiv1106.1706S,2014PASJ...66R...1T,2015arXiv150303757S,2013LRR....16....6A}. The process of extracting maximal information about new physics from these surveys will require concerted theoretical interpretation, particularly beyond the linear regime. In particular, understanding theoretical sources of uncertainty in measuring properties of the LSS will be crucial for obtaining precision constraints on new physics \cite{2016arXiv160200674B}.

The simplest statistical measure of LSS is the two-point correlation function of the density perturbation $\delta$, or its Fourier transform, the power spectrum $P(k)$, defined as
\beq \label{eq:power}
\avg{\delta(\vec{k}) \delta(\vec{k}')} =  (2\pi)^3 \delta_D(\vec{k} + \vec{k}') P(k), 
\eeq 
where the power spectrum depends only on the magnitude $ k = |\vec{k}|$. Given a survey with volume $V$, the power spectrum can be estimated by dividing $k$-space into shells of width $\Delta k$ and centered at $k_i$, with volume $V_{k_i} \approx 4 \pi k_i^2 \Delta k$ for $\Delta k/k_i \ll 1$, and integrating the variance over the shell:
\begin{align}
{\hat P}(k_i) \equiv  {1 \over V} \int_{V_{k_i}} {d^3 k \over V_{k_i} } \delta(\vec{k}) \delta(-\vec{k}) \,,
\end{align}  
such that the ensemble average of ${\hat P}(k_i)$ is the average of $P(k)$ over the shell.
The precision of this estimator and the correlations between different shells centered at $k_i$ and $k_j$ are in turn determined by its covariance $C(k_i, k_j)\equiv \langle {\hat P}(k_i){\hat P}(k_j)\rangle - \langle {\hat P}(k_i)\rangle\langle {\hat P}(k_j)\rangle = C^\text{G}_{ij}+C^\text{NG}_{ij}$, where the Gaussian and non-Gaussian contributions are given by
\begin{align}
C^\text{G}_{ij}&= \frac{1}{V}\frac{(2\pi)^3}{V_{k_i}} 2 P(k_i)^2  \delta_{ij},\label{eq:covG}\\
C^\text{NG}_{ij}&=\frac{1}{V} \int_{V_{k_i}} \int_{V_{k_j}} \frac{d^3 k_1 }{V_{k_i}} \frac{d^3 k_2 }{V_{k_j}} \,T(\vec{k}_1, -\vec{k}_1, \vec{k}_2, - \vec{k}_2).\label{eq:covdef}
\end{align}
Here, $\delta_{ij}$ is the Kronecker delta and $T$ is the trispectrum, the fourth-order connected moment of the density perturbation. 
In this paper we ignore effects due to a finite-sized survey window, which would generate an additional contribution to the covariance, the so-called super-sample covariance \cite{Takada:2013bfn,2014PhRvD..89h3519L}.
In the limit where the density field is Gaussian, the covariance is expected to be diagonal and completely determined by \Eq{eq:covG}. However, even with Gaussian initial conditions, gravitational interactions couple different Fourier modes and induce a non-Gaussian contribution through the trispectrum \cite{2012PhRvD..86h3504Y,2011JCAP...10..031B, 2013PhRvD..87l3504T}. At short distance scales, where much of the sensitivity of galaxy and weak-lensing surveys is, mode-coupling becomes increasingly relevant and understanding non-Gaussian correlations becomes crucial for extracting cosmological parameters~\cite{2014MNRAS.445.3382M}.

Thus far, understanding of the non-Gaussian covariance has either relied on versions of the astrophysically-motivated halo model~\cite{2013PhRvD..87l3504T,2014MNRAS.445.3382M}, or numerical simulations of structure formation which are computationally expensive because of the large number of realizations required for statistical convergence~\cite{2011ApJ...734...76S,2014PhRvD..89h3519L,2013PhRvD..88f3537D,2015MNRAS.446.1756B,2015arXiv151205383B}. 
While SPT may also be employed, it lacks, for instance, a clear prescription on how to treat modes in the non-linear regime~\cite{2009PhRvD..80d3531C}. Alternative versions of this formalism attempt to improve convergence by resumming a subclass of diagrams, by using the Lagrangian formulation of the theory (as inspired by the Zel'dovich approximation~\cite{1970A&A.....5...84Z}), or by other approximation schemes~\cite{2006PhRvD..73f3519C, 2007A&A...465..725V, 2007JCAP...06..026M, 2008ApJ...674..617T,2008PhRvD..77f3530M,2007PhRvD..75d3514M,2008PhRvD..78j3521B,2008PhRvD..78h3503B}. 
 However, many of these schemes have no theoretical control on quantifying the error of their approximations, are invalid near the onset of shell crossing, and may not obey all the relevant symmetries of the system, such as Galilean invariance~\cite{2009PhRvD..80d3531C}. Recent work has shown that, in simplifying limits, many of these formulations converge at sufficiently high order to the same prediction, but still fail to capture relevant effects from physics on smaller cosmological scales~\cite{2015arXiv150207389M}. 

In this paper, we compute the non-Gaussian covariance using the EFT of LSS~\cite{2012JCAP...07..051B,2012JHEP...09..082C,2014PhRvD..89d3521H}. The general idea of EFT is to describe the physics above a given length scale by considering all interactions compatible with symmetries, thus capturing the feedback from underlying microphysics in a model-independent way. For the case of LSS, this length scale, denoted $1/k_{\rm NL}$, is where non-linear effects become significant. The EFT encodes the physics of short-scale non-linear modes, with characteristic wavenumber $k \gtrsim k_{\rm NL}$, through including interactions involving only long-scale modes, with characteristic wavenumber $k \ll k_{\rm NL}$. This procedure makes SPT well-defined in the regime $k \ll k_{\rm NL}$, and can be systematically improved by including higher-order corrections.

We present the first calculation of the non-Gaussian covariance at one-loop order in the Eulerian framework, including contributions from both SPT and the leading EFT of LSS corrections. 
For efficient and accurate numerical evaluation of these contributions,
we have developed a package called \verb+FnFast+.
The EFT corrections depend on three coefficients, once the four propagated from lower orders (one from the power spectrum and three from the bispectrum) are set to their previously measured values. For one of the simulations we consider, in the basis where the three new operators are maximally uncorrelated, we find that two of them are suppressed at the few percent level
compared to the other EFT contributions. They are thus negligible given the precision of the data, and the remaining single parameter is extracted. 

The rest of the paper is organized as follows. In Sec.~\ref{sec:eom}, we review the equations of motion for the EFT of LSS in Eulerian space. Section~\ref{sec:cov} is dedicated to the calculation of the covariance. The general setup is described in Sec.~\ref{sec:setup}, before deriving the complete set of SPT and EFT contributions in Secs.~\ref{sec:SPT} and~\ref{sec:EFT}, respectively. In Sec.~\ref{sec:simulations}, we use N-body simulation data from Refs.~\cite{2014PhRvD..89h3519L} and~\cite{2015MNRAS.446.1756B,2015arXiv151205383B} as a comparison for the one-loop SPT prediction, as well as to extract the EFT coefficients. Details of the calculation and a description of the package \verb+FnFast+ are contained in the appendices.

\section{Equations of Motion}
\label{sec:eom}
We begin in this section with a basic review of the equations of motion in SPT and the EFT of LSS, emphasizing the role of effective operators in defining a self-consistent theory. To derive the Eulerian-space equations for a self-gravitating fluid made up of $n$ particles of mass $m$, we work in the conformal Newtonian limit, and consider moments of the Boltzmann equation,
\beq 
  \pd{f}{\uptau} + \frac{p^i}{m a } \pd{f}{x^i} - m \, a \sum_{n, n'} \pd{\phi_{n'}}{x^i} \pd{f_n}{p^i} = 0  \,.
\eeq
Taking the zeroth and first order moments of the comoving phase space distribution $f$,
\begin{align*} 
& \rho(\vec{x}, \uptau) = m \int d^3 p \, f(\uptau,\vec{x}, \vec{p})\, ,  \numberthis \\
&\pi^i(\vec{x}, \uptau) =  \int d^3 p \, p^i \, f(\uptau,\vec{x}, \vec{p}) \, , \numberthis
\end{align*}  
which correspond to the comoving mass and momentum density, we obtain the usual SPT equations of motion:
\beq\label{eq:SPT}
\begin{split}
\dot{\delta} + {\partial_i \pi^i \over {a\bar \rho} } &= 0\,, \\
\dot{\pi}^i +  a\rho \,  \partial^i \phi + \partial_j\left(\frac{\pi^i \pi^j }{a\rho }\right) &=0\,.
\end{split}
\eeq
Here, the dots denote derivatives with respect to the conformal time, $\mathcal{H}$ is the conformal Hubble parameter, $\phi$ is the gravitational potential that obeys the Poisson equation, and $\delta = \rho/{\bar \rho} - 1 $ is the density perturbation defined in terms of the mean matter density $\bar \rho$. 

Note that this system includes both short-scale non-linear ($k\gtrsim k_{\rm NL}$) and long-scale ($k \ll k_{\rm NL}$) modes, and that the former cannot be treated perturbatively. An analytic solution for the long-scale modes is possible through perturbation theory, but we must first derive equations of motion in terms of only long-scale modes, with the short-scale modes integrated out (i.e.~marginalized over).

To this end, we proceed by smoothing over the small-scale non-linear features of the fields, corresponding to modes with $k\gtrsim k_{\rm NL}$. Let us define smoothed, long-wavelength observables by convolving with a smoothing function $W_\Lambda$ with characteristic scale $\Lambda \ll k_{\rm NL}$, for instance,\footnote{One can take, for example, a Gaussian smoothing function $W_\Lambda\propto\exp\left(-\frac{1}{2}|\vec{x} - \vec{x}'|^2\Lambda^2\right)$.}
\beq 
\rho_l(\vec{x}, \uptau) \equiv \int d^3 x' \,W_\Lambda (\vec{x} - \vec{x}')\, \rho(\vec{x}', \uptau) \, .
\eeq
Upon smoothing, the system in \Eq{eq:SPT} is modified to
\beq
\label{eq:eom}
\begin{split}
\dot{\delta_l} + {\partial_i \pi_l^i \over {a\bar \rho}} &= 0\,, \\
\dot{\pi}_l^i + a\rho_l\,  \partial^i \phi_l + \partial_j\left(\frac{\pi_l^i \pi_l^j }{a\rho_l}\right) &=  - \partial_j \tau^{ij}\,,
\\ \partial^2 \phi_l &= \frac{3}{2} \Omega_m \mathcal{H}^2 \delta_l\,,
\end{split}
\eeq
where we have included the Poisson equation. This system describes the dynamics of long-scale modes, and can be consistently solved perturbatively for the fields $\delta_l$, $\pi_l$, and $\phi_l$.

Equations~(\ref{eq:SPT}) and~(\ref{eq:eom}) yield the same prediction for correlation functions in the linear regime, though they are distinctively different in the weakly non-linear regime.
While the purely perturbative treatment of Eq.~(\ref{eq:SPT}) breaks down, Eq.~(\ref{eq:eom}) is able to parametrize the feedback of the non-linear short-distance modes on the long-distance modes through the stress tensor. In practice, $\tau^{ij}$ is constructed from all possible local interactions of the long-scale modes that are compatible with symmetries, such as rotational and Galilean invariance, and is organized as an expansion in $k/k_\text{NL}$. For example, one of the leading operators is $\sim  c_s  \delta^{ij} \delta_l$, where $c_s$ has the physical interpretation of the speed of sound. The coefficients of some of the EFT operators in $\tau^{ij}$ are used to remove the unphysical sensitivity of SPT to the high-$k$ modes of loop integrals. The resemblance of this procedure to renormalization in quantum field theory will cause us to refer to the $\tau^{ij}$ operators as ``counterterms''. The coefficients of the EFT operators can be fixed, in principle, by fitting  to simulation data while still in the mildly non-linear regime. 

In the next section, we thus begin by deriving the SPT contributions to the covariance at one loop. This sets the scene for adding the $\tau^{ij}$ counterterms in the EFT of LSS that will allow us to obtain physical predictions.

\section{One-Loop Covariance}
\label{sec:cov}
In this section we describe the calculation of the one-loop non-Gaussian covariance in SPT and EFT. We show that the EFT operators consistently cancel the ultraviolet (UV) sensitivity of the integrals in the one-loop SPT calculation, and we provide a set of independent operators for the leading EFT correction to the SPT result. 

\subsection{Setup}
\label{sec:setup}
For solving the equations of motion given in \Eq{eq:eom}, it is convenient to work in terms of the velocity. The smoothed momentum can be written as
\begin{align*}
\label{eq:pi}
\pi^i_l &= \left[ a\rho v^i \right]_l \equiv a\rho_l v^i_l +a\bar{\rho}\,\Sigma^i \, . \numberthis
\end{align*}
The additional term $\Sigma^i$ arises from the smoothing of a product of fields and generates corrections to the continuity and Euler equations after substituting \Eq{eq:pi} in \Eq{eq:eom} \cite{2014JCAP...03..006M,2014PhRvD..90b3518C,2015arXiv150207389M,2015arXiv150907886A}. Since we are only interested in calculating correlators of the density perturbation $\delta_l$, we will use  a different definition of velocity, $v^i_\pi$, that reabsorbs $\Sigma^i$ \cite{2014JCAP...03..006M,2015arXiv150907886A},
\begin{align}
v^i_\pi\equiv v_l^i+\Sigma^i/(1+\delta_l)= \pi^i_l /(a\rho_l).
\end{align}
Equivalently, we could have opted not to make the field redefinition above, allowing us to consistently compute correlators of the physical velocity as well. Nonetheless, consistent with the field redefintion, the correlators for the density contrast would remain independent of $\Sigma^i$ (see \cite{Bertolini:2016bmt} for a detailed discussion of this point).

It is also convenient to decompose the velocity into its divergence and vorticity components, $\theta=\partial_iv_\pi^i$ and $\omega_i=\epsilon_{ijm}\partial^jv_\pi^m$. The equations of motion in Fourier space then read
\begin{widetext}
\begin{align*}
\dot{\delta}(\vec{k})+\theta(\vec{k})&= -\int d^3q\,\left[\alpha(\vec{q},\vec{k}-\vec{q})\,
\theta(\vec{q})\delta(\vec{k}-\vec{q})-\alpha^\omega_i(\vec{q},\vec{k}-\vec{q})\,\omega^i(\vec{q})\delta(\vec{k}-\vec{q})\right],\numberthis\label{eq:cont}\\
\dot{\theta}(\vec{k})+\mathcal{H}\theta(\vec{k})+\frac{3}{2}\mathcal{H}^2\Omega_m\delta(\vec{k})&= -\int d^3q
\left[\beta(\vec{q},\vec{k}-\vec{q})\theta({\vec{q}})\,\theta(\vec{k}-\vec{q})+\beta^\omega_i(\vec{q},\vec{k}-\vec{q})\,\omega^i(\vec{q})\theta(\vec{k}-\vec{q})\right]-\partial_i\left(\frac{\partial_j\tau^{ij}}{1+\delta}\right)+..., \numberthis\label{eq:euler}\\
\dot{\omega}_i(\vec{k})+\mathcal{H}\,\omega_i(\vec{k})&=- \epsilon_{ijm}\partial^j\left(\frac{\partial_s\tau^{ms}}{1+\delta}\right)+...,\numberthis\label{eq:omega}
\end{align*}
\end{widetext}
where the kernels are collected in Appendix~\ref{sptdiags},\footnote{Note that $\beta_i^\omega$ (see \Eq{eq:abomega}) disagrees with the one presented in Refs.~\cite{2015arXiv150907886A} and~\cite{Pueblas:2008uv}. This distinction is crucial for a consistent renormalization of the trispectrum. } 
and the ellipsis denotes additional terms that do not enter the calculation of the trispectrum at one-loop and at $\mathcal{O}(k^2/k^2_\text{NL})$ in the EFT corrections.
 We have assumed that the SPT contribution to the vorticity can be neglected, since at later times it is suppressed by additional powers of the growth factor compared to the other fields at the same order in perturbations \cite{2002PhR...367....1B}. Thus, vorticity is only sourced by the stress tensor.
To make the notation less cumbersome, we have dropped the explicit time dependence and the subscript $l$ (noting that all quantities for the rest of the paper are in terms of smoothed fields).
The linear solution is recovered when the right-hand side of \Eqss{eq:cont}{eq:euler}{eq:omega} is set to zero. As is standard, we use the perturbative ansatz for the growing modes:
\beq
\label{eq:ansatz}
\begin{split}
\delta(\vec{k},\uptau)&=\sum_{n=1}^\infty\left[ D^n(\uptau)\,\delta_n(\vec{k})+\varepsilon\,D^{n+2}(\uptau)\,\tilde{\delta}_n(\vec{k})\right],\\
\theta(\vec{k},\uptau)&=-\mathcal{H}f(\uptau)\sum_{n=1}^\infty\left[ D^n(\uptau)\,\theta_n(\vec{k})+\varepsilon\,D^{n+2}(\uptau)\,\tilde{\theta}_n(\vec{k})\right],\\
\omega^i(\vec{k},\uptau)&=-\mathcal{H}f(\uptau)\sum_{n=2}^\infty \varepsilon\,D^{n+2}(\uptau)\,\tilde{\omega}_{n}^i(\vec{k}),
\end{split}
\eeq
where each of the fields on the right-hand side can be written in terms of $n$ powers of the linear density perturbation which is small on large scales, $\delta_1\ll 1$.
The first term on the right hand side of the first two equations contains the standard SPT perturbative ansatz, while an $\varepsilon$ is introduced to track the leading EFT corrections, which are of order $\mathcal{O}(k^2/k^2_\text{NL})$.   As detailed in \Sec{sec:EFT}, the terms involving the stress tensor in \Eqs{eq:euler}{eq:omega} can also be expanded both in powers of $\delta_1$ and in powers of $\varepsilon$, starting at $\mathcal{O}(\varepsilon)$. Above, 
 $D(\uptau)$ is the linear growth function, $f(\uptau)=1/\mathcal{H}\,d\ln D(\uptau)/d\uptau$, and we assume $f(\uptau) = \sqrt{\Omega_m}$. For an EdS universe ($\Omega_m=1$) the solution can always be written in the form of \Eq{eq:ansatz}, with $D(\uptau)=a(\uptau)$ and $f(\uptau)=1$. Even though for a $\Lambda{\rm CDM}$ universe the time-dependence should be recomputed at each order in perturbations, it has been shown that \Eq{eq:ansatz}  is a good approximation. For the one-loop power spectrum and bispectrum, for example, the approximation is valid up to corrections of $\order(1\%)$~\cite{2008PThPh.120..549T,2012JHEP...09..082C,2015JCAP...05..007B}. The exponent $n+2$ for the EFT time-dependence is chosen such that the EFT contributions have the same time-dependence as the loop contributions from SPT.
 
With the ansatz in \Eq{eq:ansatz} one can solve Eqs.~(\ref{eq:cont} - \ref{eq:omega}) order by order. At each perturbative order $n$, the $\mathcal{O}(\varepsilon^0)$ equations will produce the SPT solution and the $\mathcal{O}(\varepsilon)$ will determine the leading EFT correction. Each field in \Eq{eq:ansatz} can be written as a convolution of $n$ linear density perturbations with kernels as
 \begin{align*}
\begin{pmatrix} 
\delta_n(\vec{k}) \\ 
\theta_n(\vec{k}) \\
\tilde{\delta}_n(\vec{k}) \\
\tilde{\theta}_n(\vec{k})\\
\tilde{\omega}_n^i(\vec{k})  
\end{pmatrix}& = 
\int \dbar^{\,3}q_1...\dbar^{\,3}q_n \, \begin{pmatrix} F_n(\vec{q}_1,...,\vec{q}_n) \\ G_n(\vec{q}_1,...,\vec{q}_n) \\ \widetilde{F}_n(\vec{q}_1,...,\vec{q}_n) \\
\widetilde{G}_n(\vec{q}_1,...,\vec{q}_n)\\
\widetilde{G}_{n}^{\omega i}(\vec{q}_1,...,\vec{q}_n)
\end{pmatrix} \\
&\times (2 \pi)^3 \delta_D\left(\vec{k}-\sum_{i=1}^n{\vec{q}_i}\right) \delta_1(\vec{q}_1)... \delta_1(\vec{q}_n),
 \numberthis \label{eq:kernels}\end{align*} 
where $\,\dbar^{\,3}q\equiv d^{\,3}q/(2\pi)^3$. The SPT kernels $F_n$ and $G_n$ can be determined from well-known recursion relations \cite{1986ApJ...311....6G, 1994ApJ...431..495J, 2002PhR...367....1B}; it will be our task to derive the EFT kernels in order to compute the EFT contribution to the covariance. 

We note that the EFT contribution to vorticity starts at order $n=2$.
As shown in \Eqs{eq:cont}{eq:euler}, this introduces $\mathcal{O}(\varepsilon)$ vorticity terms in the continuity and Euler equations starting from $n=3$, which is precisely the order we work to in computing the one-loop covariance. Thus our solution must also account for the vorticity, which was not the case for computing lower-order correlators in the EFT of LSS.

In the next two subsections, we will describe how we derived the one-loop SPT and EFT non-Gaussian covariance.

\subsection{SPT Covariance}
\label{sec:SPT}
The SPT contributions to the covariance have been computed previously at tree level~\cite{1999ApJ...527....1S}, and at one loop but with simplifying assumptions~\cite{1996ApJS..105...37S}. Here, we present the complete one-loop calculation.

The non-Gaussian covariance $C^{\rm NG}_{ij}$ corresponds to a particular configuration of the shell-averaged trispectrum as defined in \Eq{eq:covdef}. Using the perturbative setup described in the previous section, we define the different contributions to the trispectrum $T_{abcd}$ as
\beq
\label{eq:tri}
\begin{split}
&\langle\delta_a(\vec{k}_1)\delta_b(\vec{k}_2)\delta_c(\vec{k}_3)\delta_d(\vec{k}_4)\rangle\\
&\equiv(2\pi)^3\delta_D(\vec{k}_1+\vec{k}_2+\vec{k}_3+\vec{k}_4)\,T_{abcd}(\vec{k}_1,\vec{k}_2,\vec{k}_3,\vec{k}_4).
\end{split}
\eeq
Following the diagrammatic representation for these contributions (e.g., see Ref.~\cite{1996ApJS..105...37S}), the tree-level amplitudes correspond to $T_{2211}$ and $T_{3111}$, and one-loop to $T_{5111},T_{4211}, T_{3221}, T_{3311}$ and $T_{2222}$. The diagrams and explicit expressions for these can be found in \App{sptdiags}.

As discussed in \Sec{sec:eom}, the one-loop SPT contributions from integration over modes with wavenumber $q \gtrsim k_{\rm NL}$ are unphysical. To understand the contributions from this UV region, let us introduce a cutoff by integrating only over $q < \Lambda$, and consider the UV limit of the SPT kernels, 
\beq
\lim_{q \gg k_i} F_n (\vec{q} ,-\vec{q} , \vec{k}_1,..., \vec{k}_{n-2}) = {F_n^{(2)} \over q^2}  + {F_n^{(4)} \over q^4} + ...\,,
\eeq
where $F_n^{(m)}$ are functions of $\vec{k}_1,..., \vec{k}_{n-2}$, and the ellipsis denotes terms higher order in $1/q^2$.
This expansion allows us to classify the UV contributions according to their cutoff dependence.
As an example, the UV limit of $T_{5111}$ is given by
\begin{align*}
&T_{5111}^\text{UV}(k_1,k_2,\mu) \equiv 5! \int\dbar^{\,3}q
\left[ {F_5^{(2)}( -\vec{k}_1,\vec{k}_2, -\vec{k}_2) \over q^2} +... \right] \\
& \quad \times P_L(q) P_L^2(k_2) P_L(k_1)+(k_1 \leftrightarrow k_2),\numberthis
\end{align*}
where $\mu$ is the cosine of the angle between $\vec{k}_1$ and $\vec{k}_2$.
Including the angular average from the definition of $C_{ij}^{\rm NG}$ in  \Eq{eq:covdef}, we find
\begin{align*} 
&\langle T_{5111}^\text{UV}\rangle =
\frac{1}{2}\int d\mu\,T_{5111}^\text{UV}(k_1,k_2,\mu)\\
&=\frac{P_L(k_1)P_L^2(k_2)\sigma^2(\Lambda)}{706305600\, k_1^3 k_2^5}\Bigg[-4 k_1 k_2 (2266005\, k_1^8\\
& - 33470730\, k_1^6 k_2^2 + 187902172\, k_1^4 k_2^4 - 9879110\, k_1^2 k_2^6\\ 
&+ 1167375\, k_2^8) + 15 (k_1^2 - k_2^2)^3 (151067\, k_1^4 + 451074\, k_1^2 k_2^2\\
 &- 77825\, k_2^4) \log\frac{(k_1 + k_2)^2}{(k_1- k_2)^2}\Bigg]+(k_1\leftrightarrow k_2)+...,\numberthis\label{eq:T5111uv}
\end{align*} 
where $\sigma^2(\Lambda) \equiv 1/3 \int_{0}^\Lambda \dbar^{\,3}q P_L(q) /q^2$ contains the dependence on the cutoff $\Lambda$. 
The leading cutoff dependence of $T_{5111}^\text{UV}$ therefore scales as $k^2/k^2_\text{NL}$. The subleading terms denoted by the ellipsis in \Eq{eq:T5111uv} come from the higher order terms in the $1/q^2$ expansion, such as $F_5^{(4)}$, and scales with higher powers of $k/k_\text{NL}$.
Similarly, the leading cutoff dependence of $T_{4211}, T_{3221}$, and $T_{3311}$ scales as $k^2/k^2_\text{NL}$, and is proportional to that of the corresponding power spectrum and bispectrum diagrams.
The cutoff dependence of the other diagrams scale with higher powers of $k/k_\text{NL}$. As discussed in \Sec{sec:eom}, the stress tensor $\tau^{ij}$ provides counterterms to absorb this cutoff dependence. Hence, a physical prediction for the covariance at $\order(k^2/k^2_\text{NL})$ requires the stress tensor up to $\mathcal{O}(k^2/k^2_\text{NL})$. Treatment of the higher order cutoff dependence, such as those involving the $\mathcal{O}(k^4/k^4_\text{NL})$ stochastic terms, are beyond the scope of this work.

The full one-loop prediction of the covariance will include all the diagrams in \App{sptdiags}. For the loop integrations we choose a UV cutoff $\Lambda=10$ h/Mpc, which is much larger than the external wavenumber scales we are interested in. In addition, in order to compare to simulation data, the linear power spectrum is set to zero for wavenumbers $k<1/L$, where $L$ is the size of the box employed in the simulation.
For efficient and accurate numerical evaluation, we have developed a package called \verb+FnFast+, described in Appendix~\ref{appFnFast}. It is a fast, automated code written in {\tt C++} that accounts for internal symmetry factors and external wavenumber permutations, and systematically treats spurious divergences that appear when the loop wavenumber $\vec{q}$ is much less than the external $\vec{k}_i$, i.e., in the infrared region of loop integration. The integrals are computed via Monte Carlo sampling using the \verb+VEGAS+ algorithm in \verb+Cuba+. Integrals for all diagrams were independently checked using \verb+Mathematica+.

\subsection{Stress Tensor and EFT Covariance}
\label{sec:EFT}
Let us now turn to the EFT corrections to the trispectrum at $\order(k^2/k_{\rm NL}^2)$, defined analogous to Eq.~(\ref{eq:tri}) as
\beq\label{eq:triEFT}
\begin{split}
&\langle{\tilde \delta}_a(\vec{k}_1)\delta_b(\vec{k}_2)\delta_c(\vec{k}_3)\delta_d(\vec{k}_4)\rangle\\
&\equiv(2\pi)^3\delta_D(\vec{k}_1+\vec{k}_2+\vec{k}_3+\vec{k}_4)\,T_{{\tilde a}bcd}(\vec{k}_1,\vec{k}_2,\vec{k}_3,\vec{k}_4).
\end{split}
\eeq
The contributions are given by $T_{\tilde{3}111}$, $T_{\tilde{2}211}$, $T_{\tilde{1}311}$, and $T_{\tilde{1}221}$, and their diagrammatic representations and integral expressions are collected in Appendix~\ref{EFTgraveyard}. As discussed in Sec.~\ref{sec:setup}, the required kernels ${\widetilde F}_{1,2,3}$ are obtained perturbatively from the equations of motion, and depend on the stress tensor $\tau^{ij}$. The construction of the stress tensor is a straightforward task in principle, but requires some care to ensure a complete basis while avoiding a proliferation of redundant operators.

In general, the feedback from small scales induces interactions that are non-local in time. The intuitive explanation for this is that by integrating out short length scales, we are not automatically integrating out short time scales, since the linear equations of motion are scale invariant. Thus, there are memory effects present in the EFT of LSS (and also in other formalisms such as RPT, see e.g.~Ref.~\cite{2006PhRvD..73f3520C}), and these can in principle generate additional operators at sufficiently high order~\cite{2014JCAP...07..057C, 2014PhRvD..90b3518C}. 
Nonetheless, for the covariance at $\order(k^2/k_{\rm NL}^2)$, upon expanding the non-local-in-time operators order by order in the velocity, we find that the resulting set of local operators is equivalent to that obtained from a local-in-time stress tensor. They are thus physically equivalent, and we simplify the discussion here by considering the time-local case. We leave the details of time non-locality to future work~\cite{Bertolini:2016bmt}.

We consider all operators in $\partial_i \tau^{ij}$ up to $\order(\delta_1^3)$, composed from all possible contractions of the Galilean invariant building blocks ${ \partial^a \partial^b \over \partial^2} \delta$, ${ \partial^a \partial^b \over \partial^2} \theta$, and one derivative. 
For the case at hand, other building blocks, such as those using the convective derivative, do not lead to linearly independent operators. 
In Fourier space, $\partial_i \tau^{ij}$ thus takes the form 
\begin{align*}\label{eq:stresstensor}
&k_i \tau^{ij}  = \, \bar{c}_s^\delta k^j  \delta({\vec k}) + {\bar{c}_s^\theta\over \mathcal{H}f}   k^j \theta({\vec k})\\
&\quad + \int d^3q \sum_{n=1}^4 \Big[\bar{c}_n^{\delta\delta}  \delta({{\vec q}})\delta(\vec{k} - {{\vec q}}) +{\bar{c}_n^{\theta\theta}\over \mathcal{H}^2f^2}  \theta({{\vec q}})\theta(\vec{k} - {{\vec q}})\\
&\quad  + {\bar{c}_n^{\delta\theta}\over \mathcal{H}f}   \delta({{\vec q}}) \theta(\vec{k} - {{\vec q}}) 
+ {\bar{c}_n^{\theta \delta}\over \mathcal{H}f}  \theta({{\vec q}}) \delta(\vec{k} - {{\vec q}} )
 \Big]  k_i e^{ij}_n (\vec{q}, \vec{k} - \vec{q}) 
\\
&\quad  +  \int d^3q_1 d^3q_2  \sum_{n=1}^{10} \bar{c}_n^{\delta\delta\delta}   \delta({{\vec q}_1}) \delta({{\vec q}_2}) \delta(\vec{k} - {{\vec q}_1} - \vec{q}_2)  \\
&\qquad  \times k_i E^{ij}_n (\vec{q}_1, \vec{q}_2, \vec{k} - {{\vec q}_1} - \vec{q}_2) \,, \numberthis 
\end{align*} 
where the functions $e_n^{ij}$ and $E_n^{ij}$ are collected in Appendix~\ref{EFTgraveyard}. Note that with the additional derivative acting on $\partial_i \tau^{ij}$ in Eqs.~(\ref{eq:euler}) and~(\ref{eq:omega}), these operators yield the $\order(k^2/k_{\rm NL}^2)$ corrections to the covariance.
For each operator above, we have introduced a coefficient $\bar{c}$ with dimensions $[k]^{-2}$ and time dependence $\bar{c}= [{\cal H}f(\uptau)D(\uptau)]^2 c$, where $c$ is time independent. This time scaling is chosen to match the time scaling of one-loop SPT contributions.

For an efficient analysis, we must identify a corresponding minimal set of independent operators. For instance, since we are interested only in operators that are up to third order in perturbations, it follows from the degeneracy of the leading order solution, $\delta_1 = \theta_1$, that operators such as $\sim \delta \delta \theta$, etc., can be absorbed through a redefinition of the coefficients $c_n^{\delta \delta \delta}$. Similarly, a number of redundancies follow from linear dependence of the functions in Eq.~(\ref{eq:stresstensor}) once symmetrization of their arguments is considered, and once the full object ${1 \over (1 + \delta)} \partial_i \tau^{ij} $ is constructed. A strategy for paring down to a linearly independent set is to compute the solutions ${\tilde \delta}_{1,2,3}$, or equivalently the kernels ${\widetilde F}_{1,2,3}$, and then sort the resulting $k$-dependence into linearly independent functions. We find that ${\widetilde F}_1$, ${\widetilde F}_2$, and ${\widetilde F}_3$ have one, three, and eight independent operators, which can be chosen as those corresponding to $c_s^\delta\,, c_{1,2,3}^{\delta \delta}\,, c_{2,3}^{\theta \theta}\,, c_{1,2,3,4,5,6}^{\delta \delta \delta}$.

While the above counting incorporates constraints from the equations of motion, we can go further by eliminating redundancies that appear upon evaluating the correlation functions that contribute to the covariance. In particular, the angular averaging and specific $k$-configurations in the definition of $C_{ij}^{\rm NG}$ reduces the number of operators in ${\widetilde F}_3$ from eight to three. 
Thus, the effective theory prediction for the covariance involves seven operators, which can be chosen as those corresponding to $c_s^\delta\,, c_{1,2,3}^{\delta \delta}\,, c_{2,3}^{\theta \theta}\,, c_{1}^{\delta \delta \delta}$; the rest of the coefficients are set to zero. We may further define
\begin{align}
\left( \begin{array}{c}
c_s \\[0.2em]
c_1 \\[0.2em]
c_2 \\[0.2em]
c_3 \\[0.2em]
c_4 \\[0.2em]
c_5  \\[0.2em]
c_6
\end{array}\right)
\equiv
\left( \begin{array}{c}
c_s^{\delta } \\[0.2em]
c_1^{\delta \delta} \\[0.2em]
c_2^{\delta \delta}+ c_2^{\theta \theta}  \\[0.2em]
c_3^{\delta \delta}+ c_3^{\theta \theta} \\[0.2em]
c_2^{\theta \theta}+ \frac52 ( c_3^{\delta \delta}+ c_3^{\theta \theta})  \\[0.2em]
c_3^{\theta \theta}- \frac52  ( c_3^{\delta \delta}+ c_3^{\theta \theta})  \\[0.2em]
c_1^{\delta \delta \delta} + {2062 \over 2079}c_1^{\delta \delta} + {14 \over 1485}( c_3^{\delta \delta}+ c_3^{\theta \theta})
\end{array}\right) \,, 
\end{align}
such that ${\widetilde F}_1$ and ${\widetilde F}_2$ depend only on $c_s$ and $c_{1,2,3}$, while ${\widetilde F}_3$ involves the new coefficients $c_{4,5,6}$.

Having determined the stress tensor with the minimal set of operators, we may now solve Eqs.~(\ref{eq:cont} - \ref{eq:omega}) for the the kernels ${\widetilde F}_{1,2,3}$, and then evaluate the EFT corrections. We collect the results for ${\widetilde F}_{1,2,3}$ in Appendix~\ref{EFTgraveyard}. As discussed in the previous section, the EFT contributions serve as counterterms for the SPT loops. 
Indeed, we find that the function $T_{{\tilde 3}111}$ exactly matches the $\order(k^2/k_{\rm NL}^2)$ UV contribution of $T_{5111}$.
In particular,
\begin{align*}
& \langle T_{{\tilde3}111} \rangle  = \frac12 \int d \mu \, T_{{\tilde 3}111}(k_1,k_2,\mu) \\
& = {P_L(k_1)P_L^2(k_2) \over 1081080 k_1^3 k_2^5} \Bigg[ 4 k_1 k_2 \bigg(k_1^4 k_2^4 (13563 c_4+8118 c_5\\
&-124740 c_6-65607 c^\prime+31808 c_s)+5 k_1^6 k_2^2 (297 c_4\\
&-792 c_5-662 c^\prime+3544 c_s)-5 k_1^2 k_2^6 (1089 c_4+792 c_5\\
&-2894 c^\prime+980 c_s) +165 k_2^8 (9 c_4+9 c_5-26 c^\prime+20 c_s)\\
&+15 k_1^8 (99 c_5+7 c^\prime-40 c_s)\bigg)+15 (k_1-k_2)^3 (k_1+k_2)^3 \\
&\times \bigg(11 k_2^4 (9 c_4+9 c_5-26 c^\prime+20 c_s)+k_1^2 k_2^2 (-99 c_4\\
&+202 c^\prime+260 c_s)+k_1^4 (-99 c_5-7 c^\prime+40 c_s)\bigg) \\
&\times \log \frac{(k_1+k_2)^2}{(k_1-k_2)^2} \Bigg] +(k_1\leftrightarrow k_2) \,, \numberthis
\end{align*}
where $c^\prime = c_2+ c_3$. This matches the SPT result in Eq.~(\ref{eq:T5111uv}) upon setting 
\begin{align}
\left( \begin{array}{c}
c_s \\[0.35em]
c^\prime  \\[0.35em]
c_4  \\[0.35em]
c_5 \\[0.35em]
c_6 
\end{array}\right)
=
\sigma^2(\Lambda) \left( \begin{array}{c}
-{183 \over 70} \\[0.35em]
-{20991 \over 3430} \\[0.35em]
-{934103 \over 75460}  \\[0.35em]
 {22147 \over 12936} \\[0.35em]
 {5032801 \over 5093550} 
\end{array}\right) \, .
\end{align}
The cutoff dependences of $c_{s}$ and $c^\prime$ shown above are consistent with the renormalization of the power spectrum and bispectrum at $\order(k^2/k_{\rm NL}^2)$. This is a nontrivial check of the EFT, requiring a consistency between the complete set of operators derived from symmetries, the perturbative solutions ${\tilde \delta}_n$, and effects such as vorticity.\footnote{The UV matching for the full trispectrum yields $\{ c_1, c_2,c_3\} = \sigma^2(\Lambda) \{6077/6860,   - 979/245,  - 1457/686 \}$, consistent with the bispectrum.} 

In the next section, we find that two linear combinations of the three new operators yield subdominant contributions. We extract the remaining coefficient from N-body simulations of the covariance. 


\section{Extracting the EFT Coefficients from Simulations}
\label{sec:simulations}
\begin{table}
\caption{Cosmological parameters and volumes of the Li et al.~\cite{2014PhRvD..89h3519L} and Blot et al.~\cite{2015MNRAS.446.1756B, 2015arXiv151205383B} simulations.\label{tab:cosmologies} }
\begin{tabular}{c c c} 
\toprule
&Li et al.&Blot et al.\\
\midrule
$\Omega_m$ \quad& 0.286&0.257\\
$\Omega_b$ \quad& 0.047&0.044\\
h \quad&0.7&0.72\\
$n_s$ \quad&0.96&0.963\\
$\sigma_8$ \quad&0.82&0.801\\
\midrule
$V$ \quad& $(500\,\text{Mpc/h})^3$&$(656\,\text{Mpc/h})^3$\\
\bottomrule
\end{tabular}
\end{table}

We now aim to extract the new EFT coefficients from the covariance at redshift $z=0$, employing simulation data from Li et al.~\cite{2014PhRvD..89h3519L} and Blot et al.~\cite{2015MNRAS.446.1756B,2015arXiv151205383B}, whose cosmological parameters and volumes are summarized in Table~\ref{tab:cosmologies}. From the simulation data, we can in principle extract the time-dependent coefficients ${\bar c}_4$, ${\bar c}_5$, and ${\bar c}_6$, where the bar notation is defined below Eq.~(\ref{eq:stresstensor}). Consistent with our theoretical predictions, we employ datasets that do not include effects due to the survey window.
Moreover, while we include the angular average according to \Eq{eq:covdef}, we neglect the average over the bin size as the contributions to the trispectrum are slowly varying in $k$-space.

In the following subsections, we separately discuss the two datasets and their distinct implications for understanding the covariance in SPT and in the EFT of LSS.
\subsection{Fits to Li et al. Data}
\begin{figure*}[htb]
\begin{center}
\includegraphics[width=0.47\textwidth]{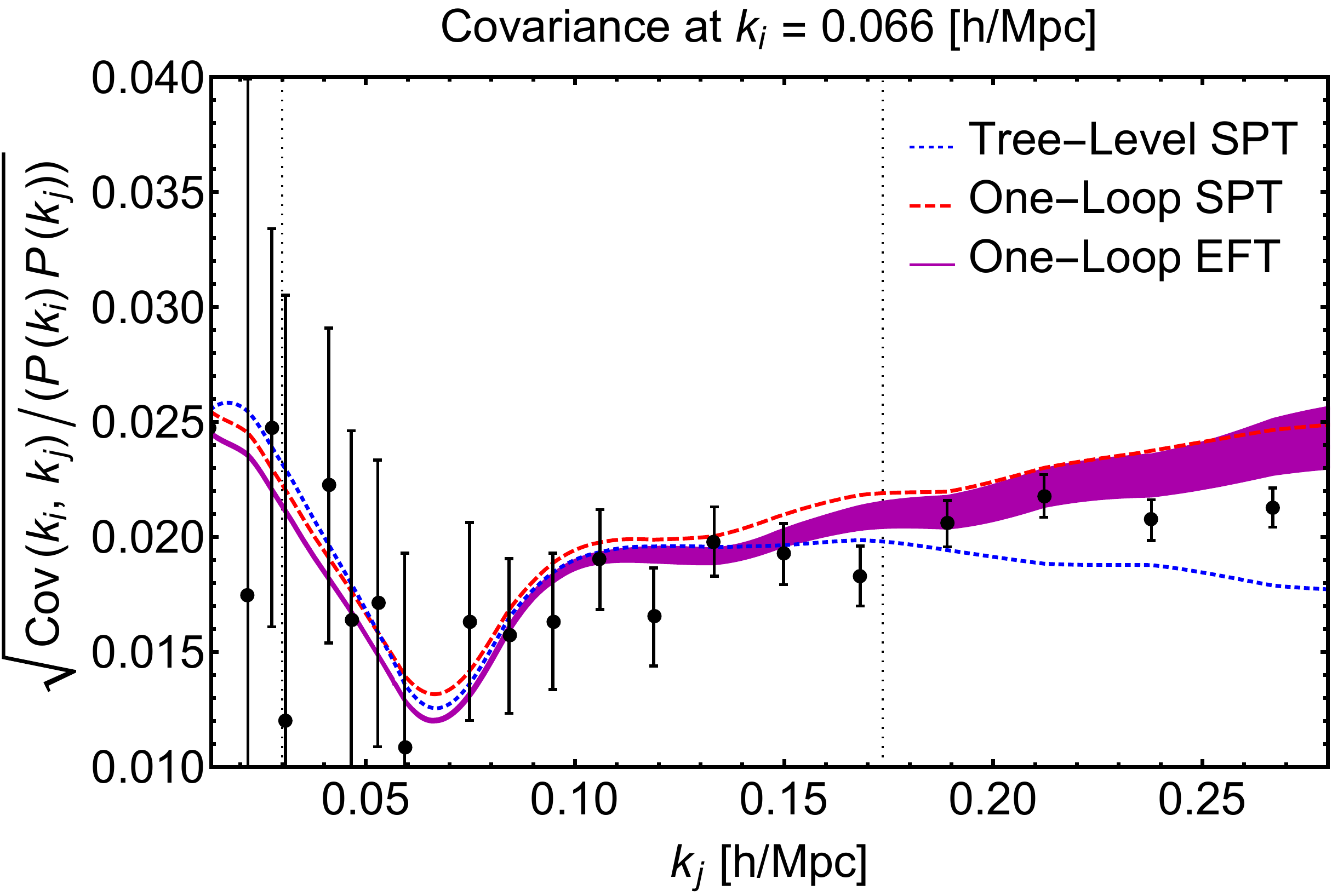}\includegraphics[width=0.47\textwidth]{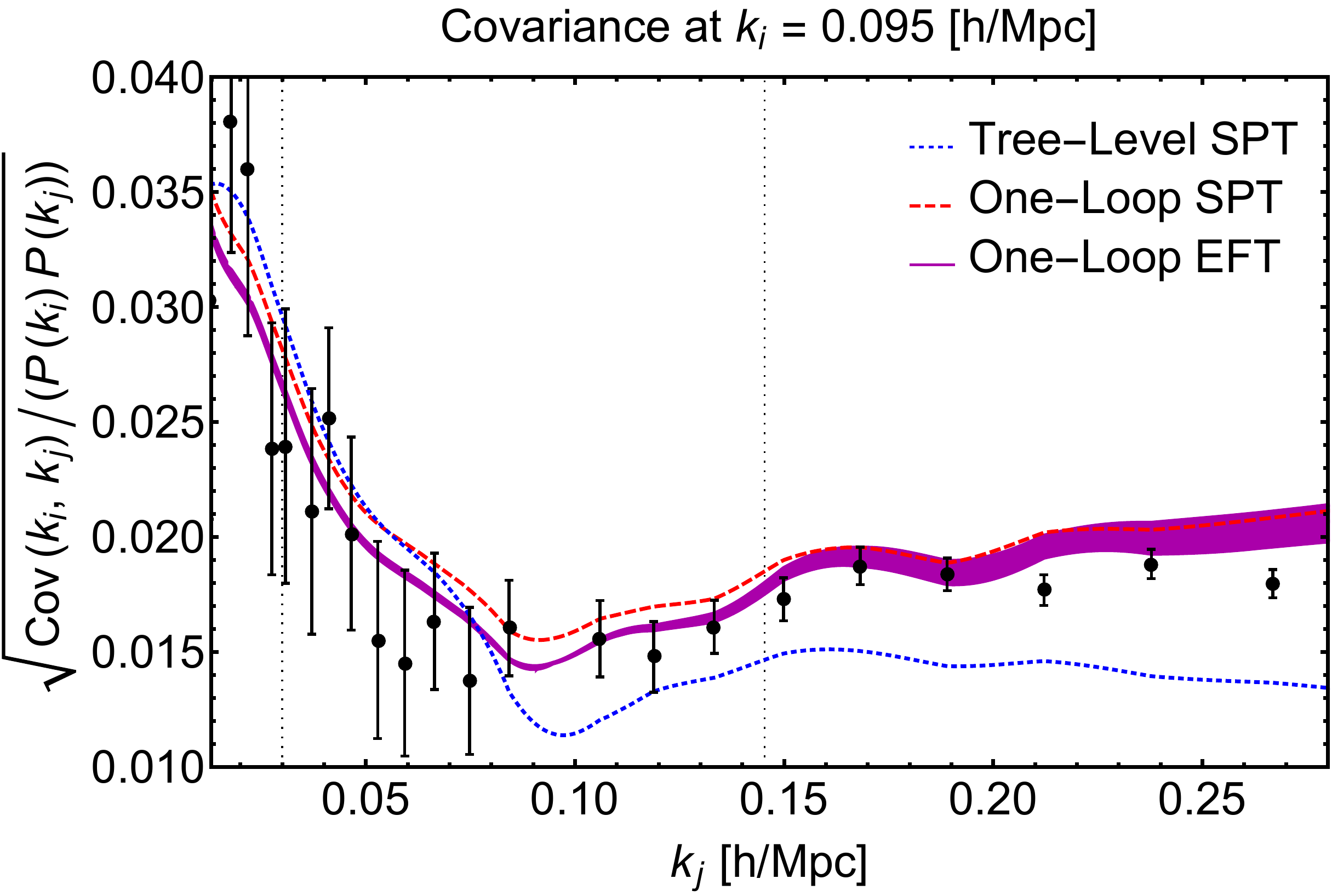}\\
\includegraphics[width=0.47\textwidth]{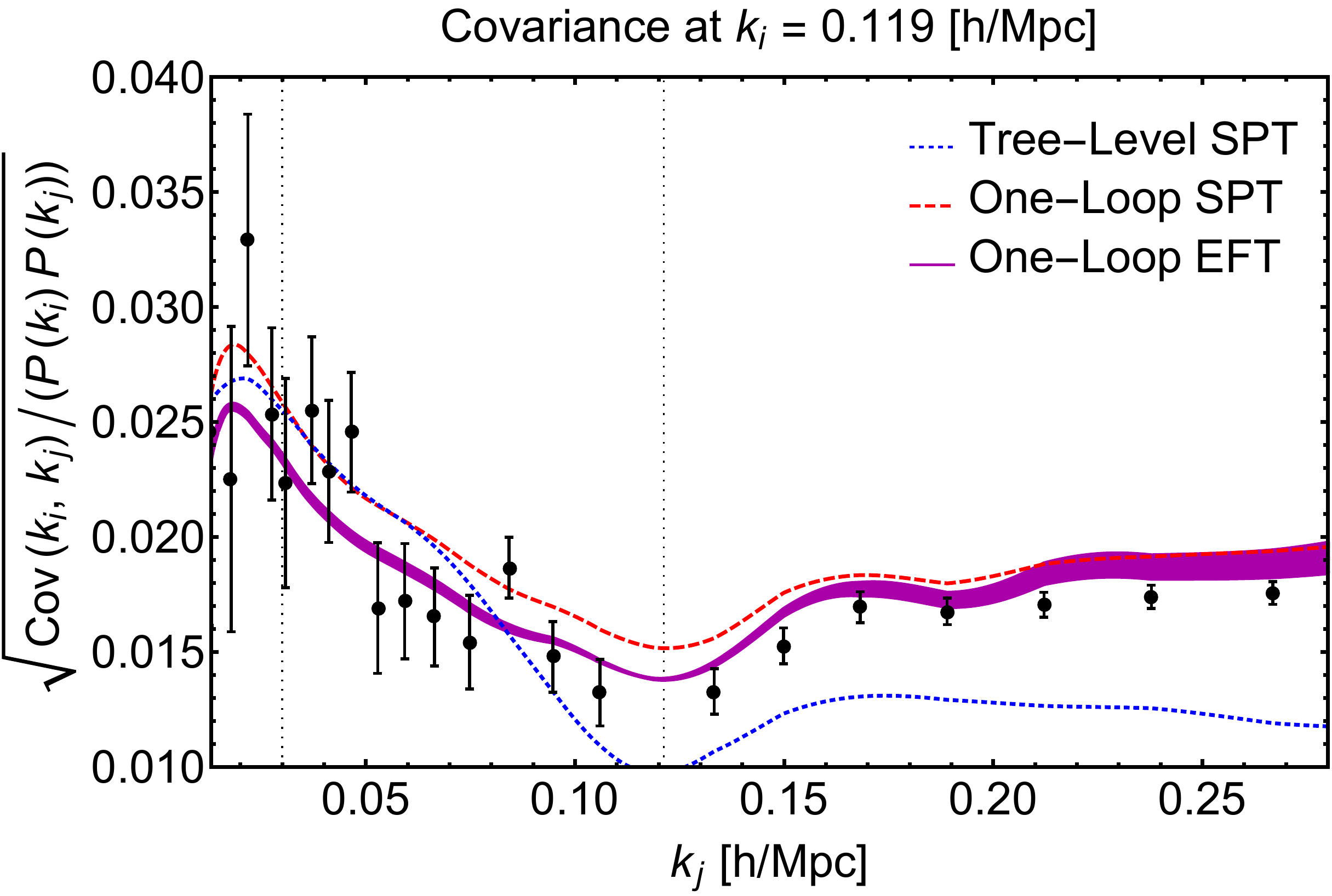}\includegraphics[width=0.47\textwidth]{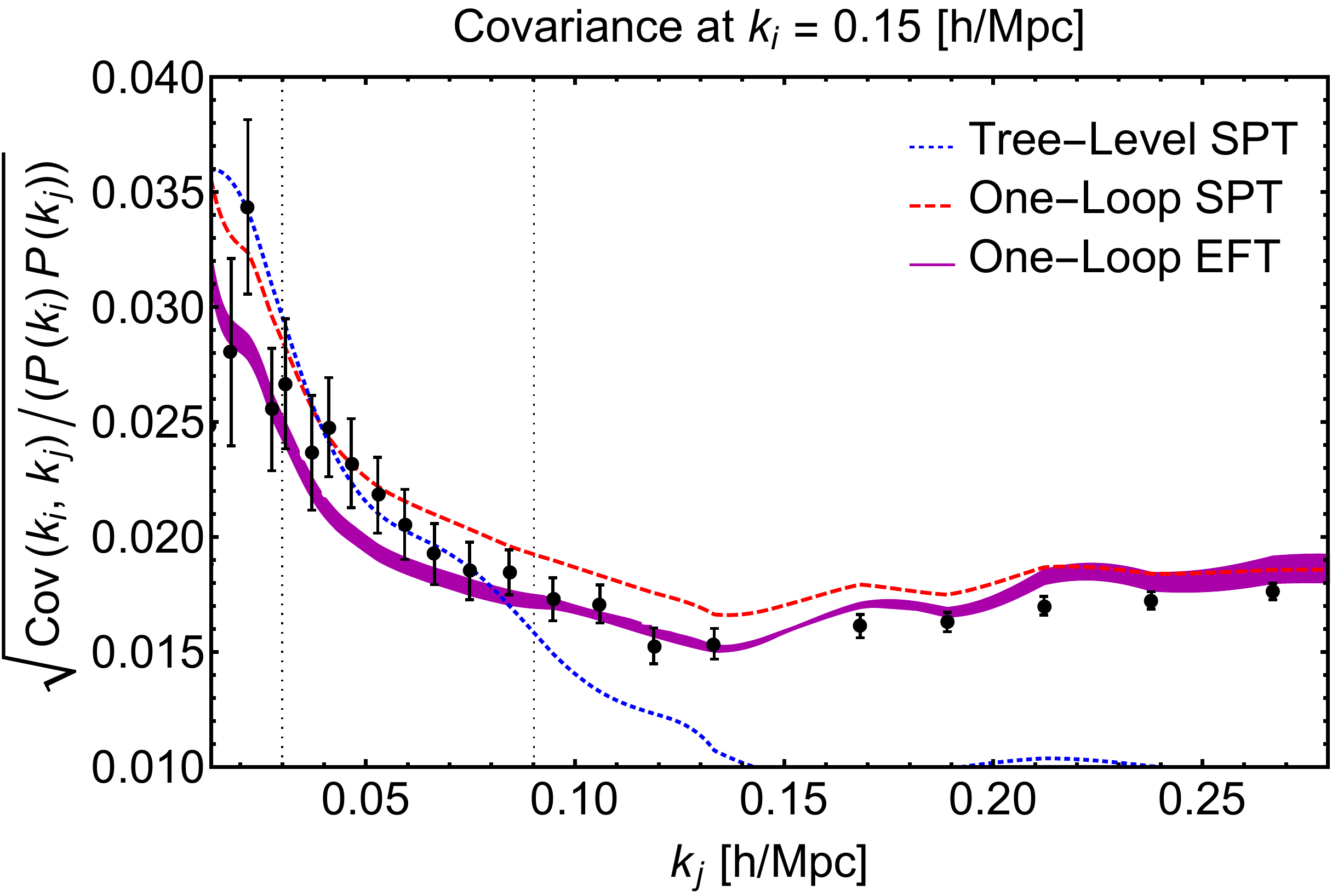}
\vspace{-0.2cm}
\caption{Several representative slices through $k$-space showing our results for the Li et al.~simulation covariance data. 
We show the covariance normalized to the power spectra, but fits are performed to the pure covariance data. Moreover, the Gaussian part of the covariance has been subtracted. The one-loop SPT contributions are evaluated with the cutoff $\Lambda =10$ h/Mpc. The thickness of the solid purple line represents the uncertainty from the fit parameter $c_*$, and from  varying the power spectrum and bispectrum coefficients by $50\%$.
Vertical dotted lines bound the region of $k_j$ values that were included in the fit for a slice at $k_i$. Any agreement between the EFT and the data to the right of the vertical lines is \emph{not} coming from fitting but from the EFT prediction with the measured parameters.} \vspace{-0.5cm}
\label{li}
\end{center}
\end{figure*}

For the cosmological parameters of the Li et al. simulation (see Table~\ref{tab:cosmologies}), we compute the SPT covariance up to one-loop order with {\tt FnFast}, using linear power spectra generated by CAMB~\cite{Lewis:1999bs}. Since the EFT operators scale as $k^2/k^2_\text{NL}$, we would ideally measure the coefficients by fitting to covariance data in the regime $k_i,k_j\ll k_\text{NL}$, such that higher-order corrections are suppressed. We find that the SPT tree-level and one-loop contributions become comparable when $k_i+k_j\sim 0.3 $ h/Mpc, and thus we estimate $k_\text{NL}\sim 0.3 $ h/Mpc. The datasets, however, have large uncertainties for $k_i, k_j\ll 0.3$ h/Mpc due to cosmic variance, and potentially unknown systematic errors. This severely limits our capacity to extract the coefficients in that region, and we are forced to include data in the range $0.03 \lesssim k_i + k_j \lesssim 0.24$ h/Mpc to get a statistically meaningful estimate of the coefficients. 

The upper cutoff of our fitting window is chosen such that $\chi^2_\nu\rightarrow1$, where $\chi^2_\nu$ is the chi squared per degree of freedom, and such that the EFT coefficients show statistical convergence. In other words, the value of the EFT coefficient is insensitive to changes in the upper cutoff of the fitting window to within our reported errors. The lower cutoff is chosen to avoid including data with large uncertainties. Additionally, beyond the sample-variance errors, we suspect that there may be large systematics that are unaccounted for in the very low-$k$ Li et al. data. For instance, there are large deviations from the trend expected from SPT, and in some cases, the covariance data is even negative, which is unphysical. Nonetheless, this excluded region represents a small number of data points, and our EFT results closely reproduce SPT in the small-$k$ limit (at the $\mathcal{O}(0.1\%)$ level near $k\sim 0.01$ h/Mpc).
We emphasize that it is likely that fitting the leading EFT corrections with this extended range leads to an overestimate of the coefficients due to saturation of signal from higher order corrections. Indeed, recent investigations have shown that early measurements of the coefficient $c_{s}$ may have been overestimated by $\sim 50\%$ (see, e.g., Refs.~\cite{2015arXiv150507098B,2015arXiv150702256B}). 

For our analysis of the EFT covariance, we use the lower-order coefficients measured from the power spectrum and bispectrum in Ref.~\cite{2015JCAP...05..007B}, which employed cosmological parameters that differ by $\order(1\%)$ to those in Table~\ref{tab:cosmologies}, and were also extracted by fitting to data up to wavenumbers $\sim 0.22$ h/Mpc.
Their coefficients $\gamma$, $\epsilon_1$, $\epsilon_2$ and $\epsilon_3$ can be mapped to ours as 
\begin{align}
\left( \begin{array}{c}
{\bar c}_s \\[0.35em]
{\bar c}_1 \\[0.35em]
{\bar c}_2  \\[0.35em]
{\bar c}_3  \\[0.35em]
\end{array}\right)
=
\left( \begin{array}{c}
9\,\gamma \\[0.35em]
\frac{9}{2}\,\gamma+\frac{33}{2}\,\epsilon_1+11\,\epsilon_2-11\,\epsilon_3 \\[0.35em]
-33\,\epsilon_2+\frac{33}{2}\,\epsilon_3 \\[0.35em]
33\,\epsilon_2  \\[0.35em]
\end{array}\right)\,.
\end{align}
The best fit values are given by
\begin{align*} 
&{\bar c}_s=13.5, &{\bar c}_1=18.5,\quad &{\bar c}_2=-41.1, &{\bar c}_3=62.4,\numberthis\label{eq:variations}
\end{align*} 
in units of ${\rm Mpc}^2/{\rm h}^2$. These inputs fix the EFT contributions $T_{\tilde{2}211}$, $T_{\tilde{1}221}$, and $T_{\tilde{1}311}$, while $T_{\tilde{3}111}$ involves the three free parameters we fit for. We account for the uncertainty on the coefficients in \Eq{eq:variations} due, e.g., to the overestimation mentioned above, by varying them by $50\%$.
 
Since the fitting forms are linear in the coefficients, we use the standard least squares formula for the estimator of the coefficients 
$\hat{c} = (G^T N^{-1} G)^{-1} G^T N^{-1} y$, where $G$ is a $n \times 3$ matrix of the 3 operators evaluated at the $n$ points we sample in $k$-space, and $N$ is the $n \times n$ covariance of the vector $y$ of $n$ data points. Note that $N$ is diagonal because we do not know the covariance of the covariance data.
For the Li et al.~data, the relevant uncertainties were estimated using a bootstrapping procedure.

For an optimal extraction of the parameters, we move to the basis of maximally uncorrelated operators, determined, e.g., by diagonalizing the covariance matrix $(G^T N^{-1} G)^{-1}$ of the estimated parameters ${\bar c}_4$, ${\bar c}_5$, and ${\bar c}_6$. We find that two operators are suppressed by $\mathcal{O}(10^{-2})$ relative to the other EFT contributions, in the entire $k_i,k_j$ domain (for $k_i \ll k_j$, the hierarchy can be traced to $k_i / k_j$ scalings). Given the $\gtrsim 10\%$ precision of the dataset, we thus neglect the suppressed operators, and obtain a prediction for the covariance with a single parameter, corresponding to the linear combination
\beq
{\bar c}_* \simeq -0.14\, {\bar c}_4 -0.04\, {\bar c}_5+ 0.99\, {\bar c}_6\, .
\eeq

Our results for the Li et al.~dataset is shown in Fig.~\ref{li}.
In the fitting region $0.03<k_i + k_j<\,$0.24 h/Mpc, we find a $\chi^2_\nu$ of 1.02 for the one-parameter EFT fit, which indicates a good fit to these data.
For SPT on the other hand, the $\chi^2_\nu$ for the Li 
data is 1.37, corresponding to a statistically significant $p$-value that is $\mathcal{O}(10^{-4})$. The best fit value of ${\bar c}_*$ is
\begin{align*}
{\bar c}_* &= 133 \pm 18 \ \ (\text{Li et al.}) \,, 
\label{cstarLi}
\numberthis
\end{align*}
in units of ${\rm Mpc}^2/{\rm h}^2$. The uncertainty includes the error obtained from the fit, as well as the effects of varying the lower-order coefficients by $50\%$. Since we expect the EFT coefficients to be $\mathcal{O}(1)$ in units of $k_\text{NL}^{-2} \sim 10 \, {\rm Mpc}^2/{\rm h}^2$, this value is reasonable given the number of $\order(1)$ factors appearing, e.g., in absorbing linearly dependent operators through redefinitions of coefficients. Moreover, as discussed above, the EFT parameter may be overestimated due to fitting in a region where higher-order corrections start to become relevant. 

\subsection{Fits to Blot et al. Data} 
\begin{figure*}[htb]
\begin{center}
\includegraphics[width=0.47\textwidth]{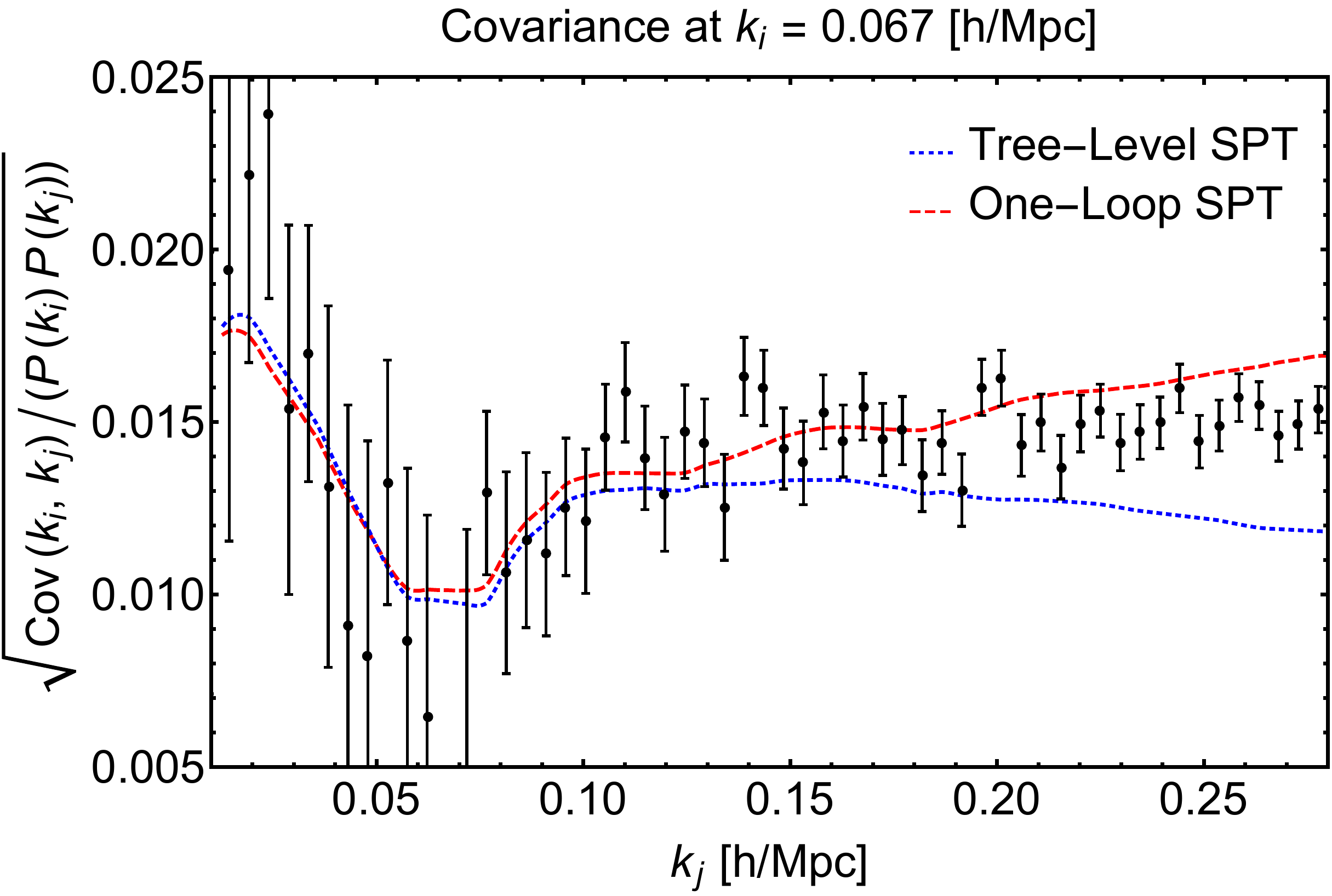}\includegraphics[width=0.47\textwidth]{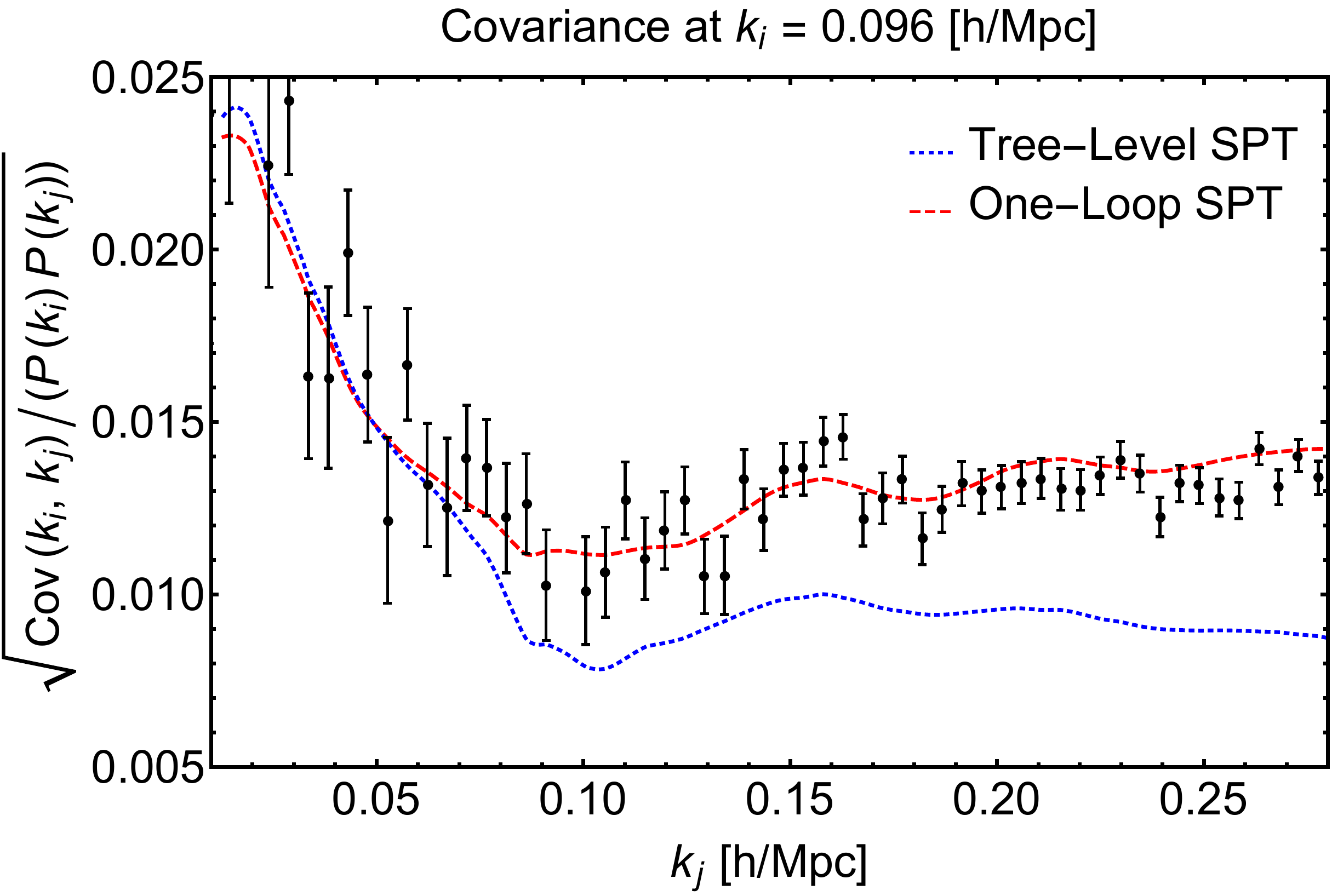}\\
\includegraphics[width=0.47\textwidth]{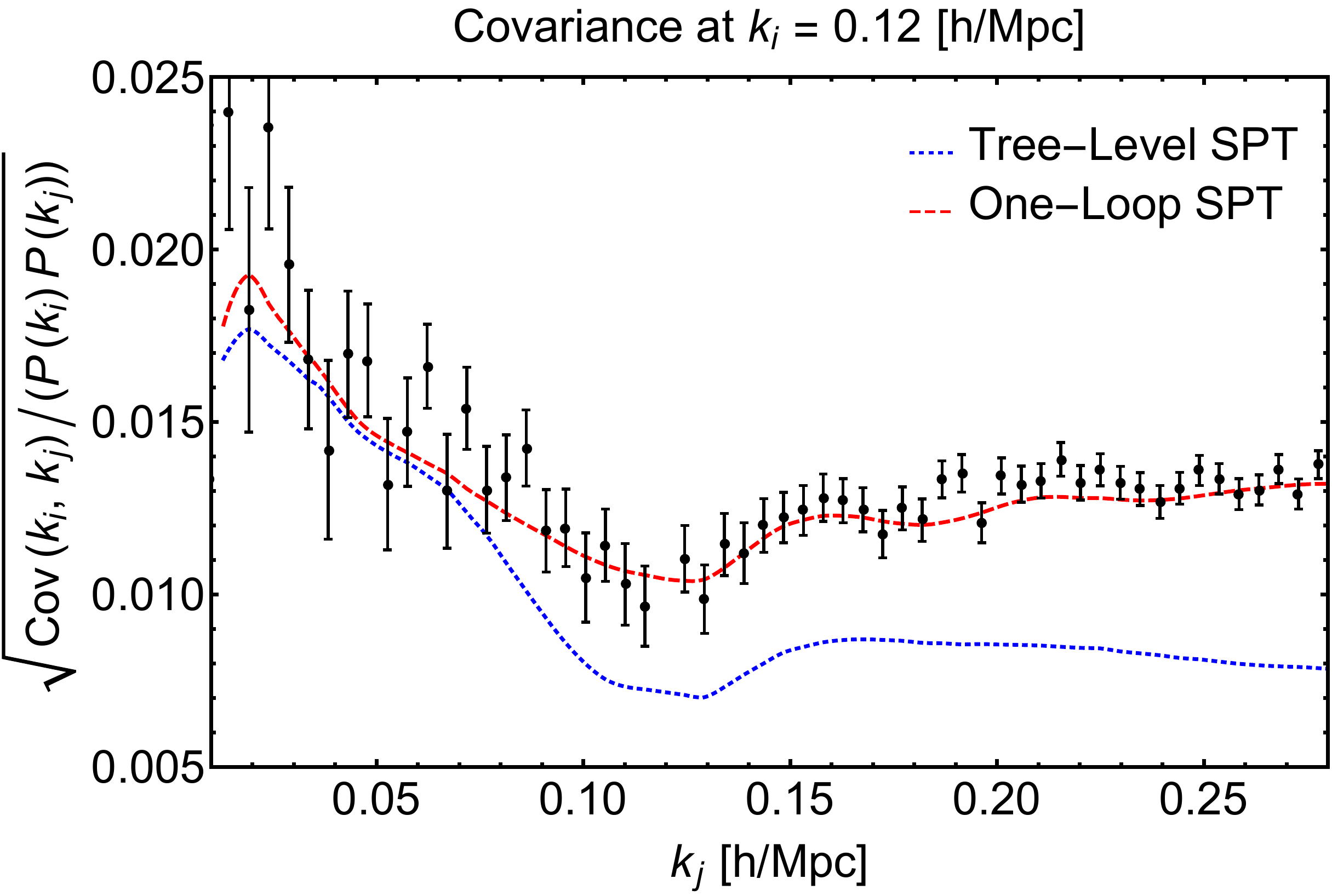}\includegraphics[width=0.47\textwidth]{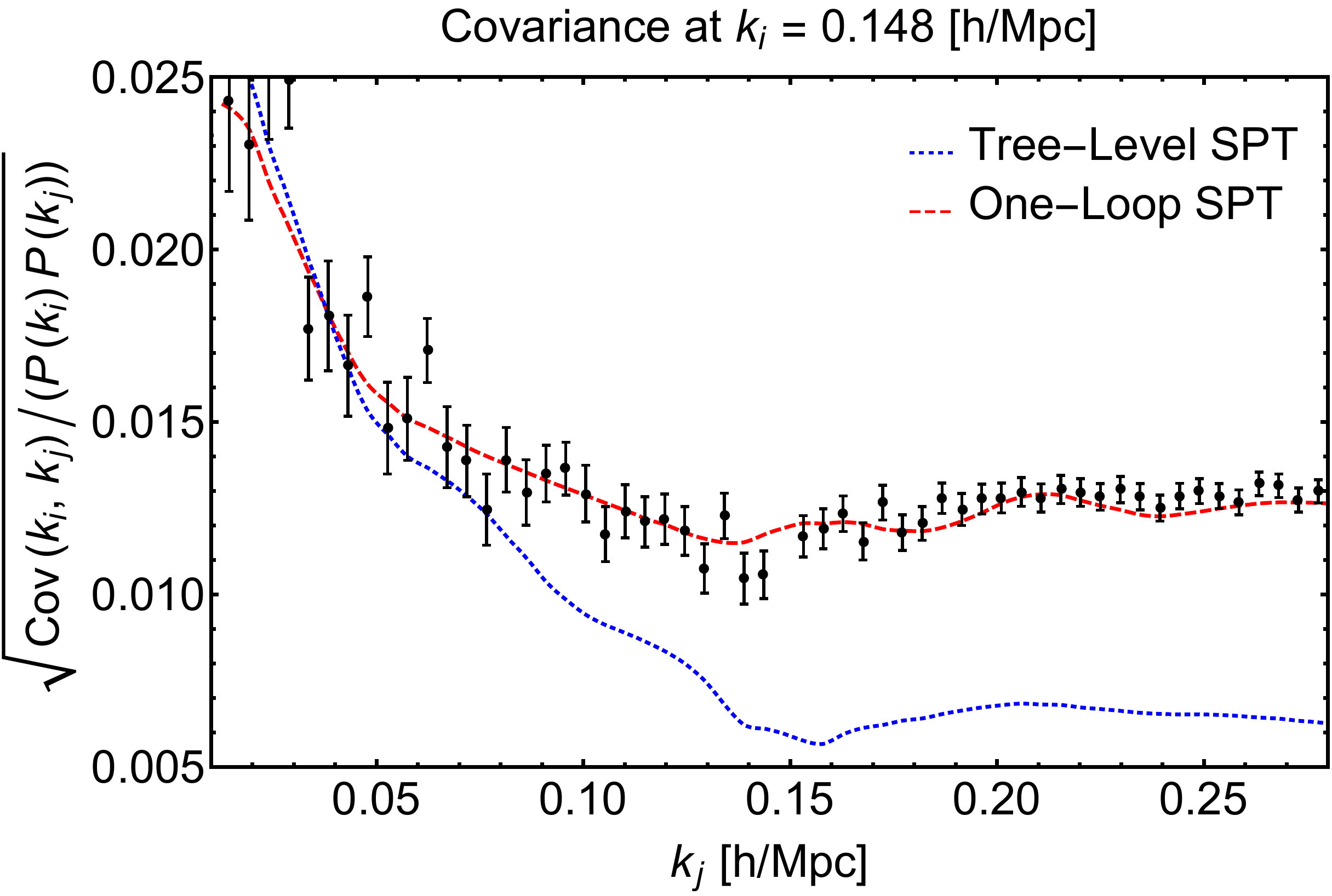}
\vspace{-0.2cm}
\caption{Same as Figure \ref{li} but with the Blot et al.~data. We find that these data are well fit by the one-loop SPT zero-parameter prediction, and there is no significant improvement from the EFT.}\vspace{-0.4cm}
\label{blot}
\end{center}
\end{figure*}

A similar procedure applied to the Blot et al.~data gives $\chi^2_\nu \rightarrow 1$ in the range $0 \lesssim k_i + k_j \lesssim 0.19$ h/Mpc. However, in that window, we obtain
\begin{align*}
{\bar c}_* &= -31 \pm 18  \ \ (\text{Blot et al.}) \,, 
\numberthis
\label{cstarBlot}
\end{align*}
in units of ${\rm Mpc}^2/{\rm h}^2$ and again including both the error from the fit and from varying the lower-order coefficients by $50\%$. Following Ref.~\cite{2015arXiv151205383B}, the errors on the covariance data are assumed to follow the Wishart distribution, and we have added an additional 10\% error to account for possible deviations in the $k$-range under study.

This value of ${\bar c}_* $, \Eq{cstarBlot}, seems inconsistent with the value obtained from the Li et al.~data, \Eq{cstarLi}.  Furthermore, unlike the Li et al.~data, the EFT does not provide a significant improvement over the SPT one-loop prediction, whose reduced chi squared $\chi^2_\nu\rightarrow 1$ is already excellent in the region $0 \lesssim k_i + k_j \lesssim 0.25$ h/Mpc.
In fact, since there are contributions from lower-order EFT counterterms with fixed coefficients (from the power spectrum and the bispectrum), we find that the remaining free parameter is optimized by the fit to compensate for these lower-order EFT contributions. In other words, the EFT is best optimized by approximately reproducing SPT in the fitting window. In \Fig{blot} we compare Blot et al.~data to the SPT predictions. Moreover, we find that for the power spectrum, the one-loop SPT prediction begins to deviate significantly from the Blot et al. data starting roughly at $k\sim 0.1$ h/Mpc, which is consistent with other studies of the one-loop SPT power spectrum. 
 
The differences of the fits highlight that there is a systematic offset between the two data sets beyond what can be accounted for by statistics or by a slight difference in cosmological parameters and volumes.  We do not seek to speculate about which of the two simulations is more accurate.  We do emphasize, however, that unfortunately the data is not currently accurate enough to distinguish between one-loop predictions from SPT and the EFT of LSS. 

There are alternate ways to measure the EFT coefficients for the covariance, which may provide further insight. For example, one could attempt to measure the trispectrum from simulations or, in principle, one could measure the same EFT covariance coefficients by performing a full measurement of EFT coefficients that are present in the two-loop power spectrum.

\section{Conclusions}\vspace{-0.2cm}
\label{sec:conclusions}
We have carried out the first one-loop calculation of the non-Gaussian covariance of the matter power spectrum, including the complete set of SPT contributions and the leading EFT of LSS corrections. Seven EFT operators, of which four appear in the power spectrum and bispectrum, parametrize the effects of non-perturbative short-scale modes on the dynamics of long-distance modes, and provide counterterms for the leading cutoff dependence of the SPT loop contributions. In a forthcoming publication, further details of this calculation will be presented in the context of the full trispectrum~\cite{Bertolini:2016bmt}.

We have also measured the coefficients of the EFT operators in the Li et al. covariance data; interestingly, we find that of the three new EFT operators that arise for the covariance, the data depends most on one particular linear combination of these operators, with the other combinations suppressed. Thus we were able to extract the coefficient of this one linear combination of operators; in spite of the fact that we are limited by the low-$k$ precision of this dataset, we find that the EFT prediction describes the data up to $k_i+k_j\sim 0.3$ h/Mpc, more than double the reach compared to SPT.  In the other simulation we considered (by Blot et al.), we found that SPT alone works well to $k_i+k_j\sim 0.25$ h/Mpc. 
We thus find that there is a systematic offset between the two datasets, which cannot be accounted for by, for instance, differences in their cosmologies.

While the EFT approach to LSS is theoretically sound and captures non-linear effects in the data, its full utility is challenged by reliable extraction of the coefficients. This points to the need for simulations with improved precision in the low-$k$ regime, or alternative ways of measuring the EFT parameters, e.g., through other observables or with input from higher-order perturbative corrections.
These developments will be important for understanding the regime of validity of the EFT, and ultimately, for building a framework for meaningful comparison between theory and data in the era of precision cosmology. 
\section*{Acknowledgements}
It is an immense pleasure to thank Simone Ferraro, Adrian Liu, Pat McDonald, Yasunori Nomura, Marcel Schmittfull, Uro\v{s} Seljak, Leonardo Senatore, Martin White, Mark Wise, and Hojin Yoo for useful conversations and correspondence pertaining to this work. We would also like to thank Linda Blot, Pier Stefano Corasaniti, and Yin Li for making their simulation data available to us and for useful discussions about the subtleties of the respective simulations. We also acknowledge the referee for his or her helpful comments on the original version of the manuscript. KS is supported by a Hertz Foundation Fellowship and by a National Science Foundation Graduate Research Fellowship. DB, MS, JRW, and KZ are supported under contract DE-AC02-05CH11231.
\appendix
\onecolumngrid
\section{SPT Kernels and Trispectrum Amplitudes}
\label{sptdiags}
In this appendix, we collect the kernels for the equations of motion, the kernels for the SPT perturbative solution, and the expressions for the one-loop SPT covariance.

The kernels appearing in Eqs.~(\ref{eq:cont} - \ref{eq:omega}) are given by
\begin{align}
\alpha(\vec{q}_1, \vec{q}_2)& \equiv \frac{ \vec{q}_1 \cdot( \vec{q}_1 +  \vec{q}_2)}{q_1^2},
&\beta( \vec{q}_1,  \vec{q}_2 ) &\equiv \frac{( \vec{q}_1 +  \vec{q}_2)^2  \vec{q}_1\cdot \vec{q}_2}{2 q_1^2 q_2^2},\\
\alpha^\omega_i(\vec{q}_1, \vec{q}_2)& \equiv \frac{(\vec{q}_2\times\vec{q}_1)_i}{q_1^2},&
\beta^\omega_i(\vec{q}_1, \vec{q}_2)& \equiv \frac{(q_2^2+2\vec{q}_1\cdot\vec{q}_2)(\vec{q}_1\times\vec{q}_2)_i}{q_1^2q_2^2}\label{eq:abomega} \,.
\end{align}

\begin{figure*}[htb]
\includegraphics[width=0.55 \textwidth]{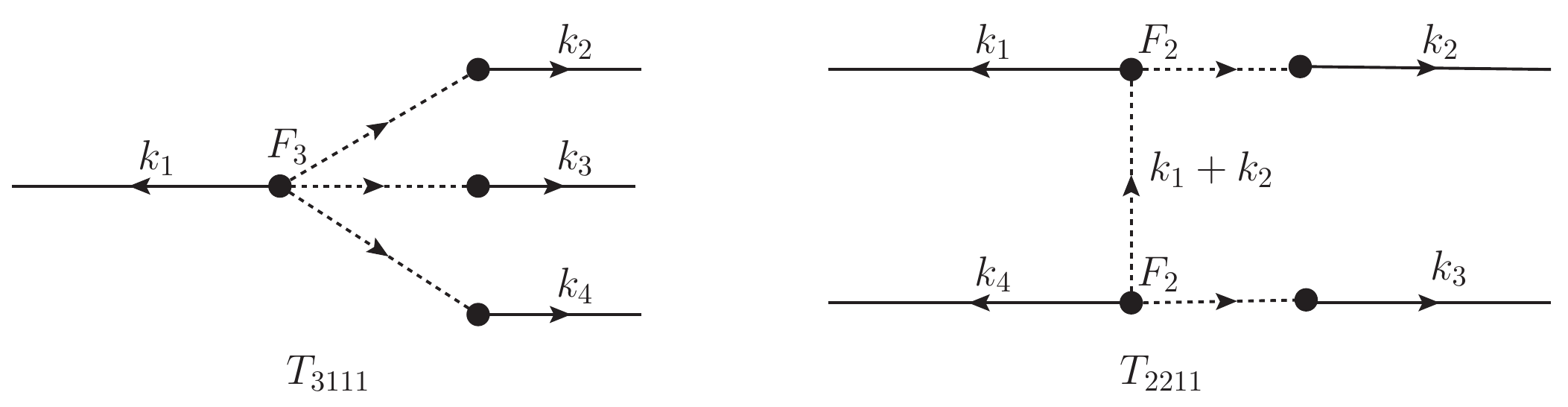}
\caption{Tree-level SPT contributions to the trispectrum, represented diagrammatically.}
\label{fig:tree}
\end{figure*}

The SPT mode-coupling kernels in \Eq{eq:kernels} obey the following recursion relations \cite{1986ApJ...311....6G, 1994ApJ...431..495J, 2002PhR...367....1B}:

\begin{align*}
F_n(\vec{q}_1,\ldots,\vec{q}_n) = \sum_{m=1}^{n-1} \frac{G_m(\vec{q}_1,\ldots,\vec{q}_m)}{(2n+3)(n-1)} \Big( &(2n+1) \alpha(\vec{k}_1, \vec{k}_2) F_{n-m}(\vec{q}_{m+1},\ldots,\vec{q} _n) + 2\beta(\vec{k}_1, \vec{k}_2) G_{n-m}(\vec{q}_{m+1},\ldots,\vec{q} _n) \Big) \\
 G_n(\vec{q}_1,\ldots,\vec{q}_n) = \sum_{m=1}^{n-1} \frac{G_m(\vec{q}_1,\ldots,\vec{q}_m)}{(2n+3)(n-1)} \Big( &3 \alpha(\vec{k}_1, \vec{k}_2) F_{n-m}(\vec{q}_{m+1},\ldots,\vec{q} _n) + 2n\beta(\vec{k}_1, \vec{k}_2) G_{n-m}(\vec{q}_{m+1},\ldots,\vec{q} _n) \Big),\numberthis \label{eq:sptKernels}\end{align*}
 where $\vec{k}_1 = \vec{q}+\ldots+\vec{q}_m$, $\vec{k}_2 = \vec{q}_{m+1}+\ldots+\vec{q}_n$, and $F_1 = G_1 = 1$. 
The kernels in \Eq{eq:sptKernels} should be further symmetrized over permutations of their arguments:
\begin{align}
F_n^{s} (\bq_{1..n}) = \frac{1}{n!} \sum_{\pi \in \sigma_n} F_n(\pi(\bq_{1..n})) \,,
\end{align}
where the sum is over the set $\sigma_n$ of permutations $\pi$ of $n$ indices.
 
Finally, we collect the expressions for the tree-level and one-loop SPT trispectrum amplitudes. We assign generic external labels $\vec{k}_{1,2,3,4}$, noting that for the covariance one has to set, e.g., $\vec{k}_2=-\vec{k}_1$ and $\vec{k}_4=-\vec{k}_3$, average over the shell, and rescale by the volume, according to \Eq{eq:covdef}. 
\begin{figure*}[htb]
\includegraphics[width=0.9 \textwidth]{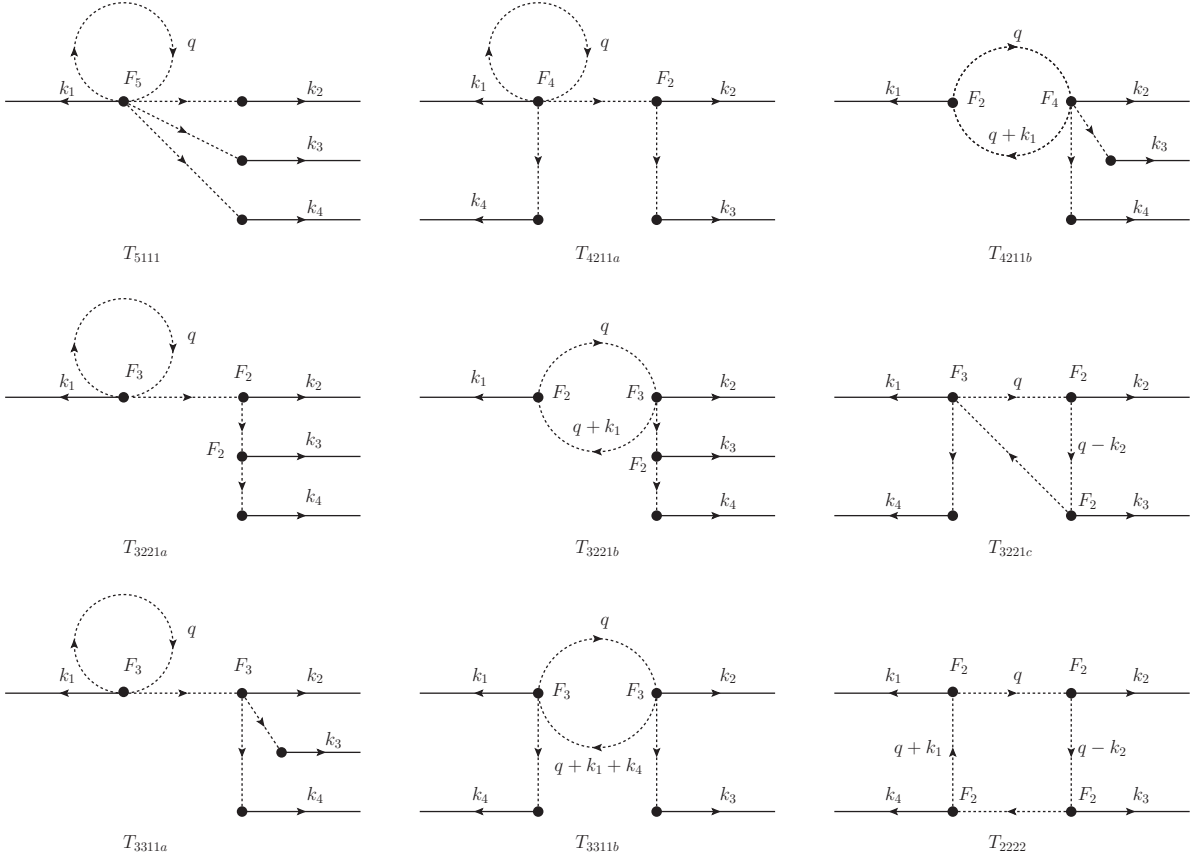}
\caption{One-loop SPT contributions to the trispectrum, represented diagrammatically.}
\label{fig:loop}
\end{figure*}
Following the diagrammatic representation for the various contributions (see e.g., Ref.\cite{1996ApJS..105...37S}), there are two tree-level SPT diagrams, which are shown in Fig.~\ref{fig:tree}. The corresponding amplitudes are given by
\begin{align*}
T_{3111} &= 3! F_3^s(\vec{k}_1,\vec{k}_2, \vec{k}_3) P_L(k_1) P_L(k_2) P_L(k_3) +3\text{ perms.},\numberthis\\
T_{2211} &= (2!)^2 F_2^s(-\vec{k}_1-\vec{k}_2, \vec{k}_2) F_2^s(\vec{k}_1+\vec{k}_2, \vec{k}_3) P_L(\abs{\vec{k}_1+\vec{k}_2}) P_L(k_2) P_L(k_3)+11\text{ perms.} \numberthis 
\end{align*}

Note that our convention is to assign positive sign to wavevectors exiting a vertex.
The one-loop SPT diagrams are shown in Fig.~\ref{fig:loop}. The amplitudes $T_{4211}$ and $T_{3311}$ receives contributions from two types of diagrams, labeled $a$ and $b$, while the amplitude $T_{3211}$ has an additional contribution labeled $c$. The four contributions with leading order cutoff dependence (i.e. with a single kernel involved in the loop integration) are
\begin{align*} 
T_{5111}& =\frac{5!}{2!} \int \dbar^{\,3}q \, F_5^s(\vec{q}, -\vec{q}, \vec{k}_2,\vec{k}_3,\vec{k}_4) P_L(q) P_L(k_2) P_L(k_3) P_L(k_4)+3\text{ perms.}, \numberthis\\
 T_{4211a} &= 4! \int \dbar^{\,3}q \, F_4^s(\vec{q}, -\vec{q}, \vec{k}_2+\vec{k}_3, \vec{k}_4) F_2^s(-\vec{k}_2-\vec{k}_3, \vec{k}_3) P_L (q) P_L(\abs{\vec{k}_2 + \vec{k}_3}) P_L(k_3) P_L(k_4)+23\text{ perms.}, \numberthis \\
 T_{3221a}& = 3!2!\int \dbar^{\,3}q \, F_3^s( \vec{q},  -\vec{q},  -\vec{k}_1) F_2^s( \vec{k}_1 ,  \vec{k}_3+ \vec{k}_4) F_2^s(-\vec{k}_3-\vec{k}_4, \vec{k}_4)\\
 &\quad\quad\quad\quad \quad\quad\quad \times P_L(q) P_L(k_1) P_L(\abs{\vec{k}_3+\vec{k}_4}) P_L(k_4)+23\text{ perms.}, \numberthis\\ 
 T_{3311a}& = \frac{(3!)^2}{2!}\int \dbar^{\,3}q \, F_3^s( \vec{q},   - \vec{q},   -\vec{k}_1) F_3^s( \vec{k}_1,  \vec{k}_3,  \vec{k}_4 ) P_L(q) P_L(k_1) P_L(k_3)P_L(k_4)+11\text{ perms.}\,\numberthis 
 \end{align*}
The five contributions with subleading cutoff dependence are given by
\begin{align*}
T_{4211b} & = 4! \int \dbar^{\,3}q \, F_4^s(\vec{k}_1 + \vec{q}, -\vec{q}, \vec{k}_3, \vec{k}_4) F_2^s(-\vec{q}-\vec{k}_1, \vec{q}) P_L(\abs{\vec{q} +\vec{k}_1})P_L(q) P_L(k_3) P_L(k_4)+11\text{ perms.},  \numberthis\\
T_{3221b}&=3!2!\int \dbar^{\,3}q \, F_3^s( \vec{q} + \vec{k}_1, -\vec{q},\vec{k}_3+\vec{k}_4) F_2^s( -\vec{k}_1 - \vec{q},\vec{q}) F_2^s(-\vec{k}_3-\vec{k}_4, \vec{k}_4)\\
&\quad\quad\quad\quad \quad\quad\quad \times P_L(q) P_L(\abs{\vec{k}_1 + \vec{q}})P_L(\abs{\vec{k}_3+\vec{k}_4}) P_L(k_4)+23\text{ perms.},\numberthis\\
T_{3221c} &= 3! (2!)^2\int \dbar^{\,3}q \, F_3^s(\vec{q}, -\vec{k}_1 - \vec{k}_4 - \vec{q}, \vec{k}_4) F_2^s(-\vec{q}, \vec{q}-\vec{k}_2) F_2^s(\vec{k}_2-\vec{q} , \vec{k}_1 + \vec{k}_4 + \vec{q})\\
&\quad\quad\quad\quad \quad\quad\quad \times P_L(q) P_L(\abs{\vec{k}_1 + \vec{k}_4 + \vec{q}}) P_L(\abs{\vec{q} - \vec{k}_2}) P_L(k_4)+11\text{ perms.},  \numberthis \\ T_{3311b} &= \frac{(3!)^2}{2!} \int \dbar^{\,3}q \, F_3^s( \vec{q},    -\vec{q}- \vec{k}_1-  \vec{k}_4,   \vec{k}_4) F_3^s( -\vec{q},    \vec{q}+ \vec{k}_1+  \vec{k}_4,   \vec{k}_3)\\
 &\quad\quad\quad\quad \quad\quad\quad \times P_L(q) P_L(\abs{\vec{q}+ \vec{k}_1+  \vec{k}_4}) P_L(k_3) P_L(k_4)+11\text{ perms.},  \numberthis\\
T_{2222} &= (2!)^4\int \dbar^{\,3}q \, F_2^s(\vec{q}, -\vec{k}_1-\vec{q}) F_2^s(-\vec{q}, \vec{q}-\vec{k}_2) F_2^s(\vec{k}_2-\vec{q},\vec{q}-\vec{k}_2-\vec{k}_3) F_2^s(\vec{k}_2+\vec{k}_3-\vec{q},\vec{q}+\vec{k}_1) \\
&\quad\quad\quad\quad \quad\quad\quad\times P_L(q) P_L(\abs{\vec{q}-\vec{k}_2}) P_L(\abs{\vec{q}-\vec{k}_2-\vec{k}_3}) P_L(\abs{\vec{q}+\vec{k}_1})+2\text{ perms.}\numberthis\\
 \end{align*}
For each amplitude, we have included a symmetry factor that accounts for the degenerate configurations of the diagram, and the number of the corresponding inequivalent permutations of external labels one has to sum over.

\section{EFT Operators, Kernels, and Counterterms}
\label{EFTgraveyard}
In this appendix we collect the functions appearing in the stress tensor in Eq.~(\ref{eq:stresstensor}), the kernels for the EFT solutions, and the EFT contributions to the covariance.

The functions $e_n^{ij}$ and $E_n^{ij}$ in Eq.~(\ref{eq:stresstensor}) are given by
\begin{align*}\label{eq:operators}
 E_1^{ij} (\vec{q}_1, \vec{q}_2, \vec{q}_3)&= e_1^{ij} (\vec{q}_1, \vec{q}_2) =\delta^{ij} \,, &
E_2^{ij} (\vec{q}_1, \vec{q}_2, \vec{q}_3)&= e_2^{ij} (\vec{q}_1, \vec{q}_2) = {q_1^i q_1^j \over q_1^2} \,, \\
E_3^{ij} (\vec{q}_1, \vec{q}_2, \vec{q}_3)&= e_3^{ij} (\vec{q}_1, \vec{q}_2) = {q_1^{\{ i} q_2^{j\}} q_1^a q_2^a \over q_1^2 q_2^2} \,, &
E_4^{ij} (\vec{q}_1, \vec{q}_2, \vec{q}_3) &= e_4^{ij} (\vec{q}_1, \vec{q}_2) = {\delta^{ij}  (q_1^a q_2^a)^2 \over q_1^2 q_2^2} \,, \\
E_5^{ij} (\vec{q}_1, \vec{q}_2, \vec{q}_3)&= {q_1^i q_1^j (q_2^a q_3^a)^2 \over q_1^2 q_2^2 q_3^2 }  \,, &
E_6^{ij} (\vec{q}_1, \vec{q}_2, \vec{q}_3)&= {q_1^{\{ i} q_2^{j\}} q_1^a q_3^a q_2^b q_3^b \over q_1^2 q_2^2 q_3^2 }\,, \\
E_7^{ij} (\vec{q}_1, \vec{q}_2, \vec{q}_3)&= {\delta^{ij} q_1^a q_2^a q_2^b q_3^b q_3^c q_1^c \over q_1^2 q_2^2 q_3^2 } \, , &
E_8^{ij} (\vec{q}_1, \vec{q}_2, \vec{q}_3)&=   { \epsilon^{\{ iab} \epsilon^{jcd \}} q_1^a q_1^c q_2^b q_2^d \over q_1^2 q_2^2  }  \,, \\ 
E_9^{ij} (\vec{q}_1, \vec{q}_2, \vec{q}_3)&=   {\epsilon^{\{ iab} \epsilon^{jcd \}} q_1^a q_1^e q_2^c q_2^e q_3^b q_3^d \over q_1^2 q_2^2 q_3^2 }\,, &
E_{10}^{ij} (\vec{q}_1, \vec{q}_2, \vec{q}_3)&=  {\delta^{ij} (\epsilon^{abc}  q_1^a q_2^b  q_3^c )^2 \over q_1^2 q_2^2 q_3^2 } \,,   \numberthis
\end{align*}
where $\{ \ \}$ denotes symmetrization in the indices $i,j$.

Given our Fourier shapes and expressions for the effective stress tensor, we can algebraically solve the equations of motion order by order for $\tilde{\delta}$ and $\tilde{\theta}$, which we express using the mode-coupling kernels (see \Eq{eq:kernels}). At leading order we find
\begin{align} &\widetilde{F}_1(\vec{k}) = -\frac{1}{9} (c_s^\delta - c_s^\theta) k^2, \\
& \widetilde{G}_1(\vec{k}) = -\frac{1}{3} ( c_s^\delta - c_s^\theta) k^2, \end{align}
where $c_s^\delta$ and $c_s^\theta$ are degenerate at this order, hence why effectively only one EFT coefficient is needed for the power spectrum.
At second order, we find
\begin{align*} &\widetilde{F}_2(\vec{k}_1, \vec{k}_2)  = \,\frac{3}{11} \alpha (\vec{k}_1, \vec{k}_2) \left( \widetilde{G}_1 (\vec{k}_1) + \tilde{F}_1(\vec{k}_2)\right) + \frac{2}{33} \beta (\vec{k}_1, \vec{k}_2) \left(\widetilde{G}_1 (\vec{k}_1) + \widetilde{G}_1 (\vec{k}_2)\right)\\& -\frac{2}{33} \,c_s^\delta\, k^2 F_2(\vec{k}_1, \vec{k}_2) +\frac{2}{33} \,c_s^\theta\, k^2 G_2(\vec{k}_1, \vec{k}_2) +\frac{2}{33} (c_s^\delta - c_s^\theta) \left(\vec{k}\cdot\vec{k}_2\right)  - \frac{2}{33} \sum_{n=1}^4 (c_n^{\delta\delta} - c_{n }^{\delta\theta}-c_n^{\theta\delta}+c_n^{\theta\theta})\, k_ik_j\,e_n^{ij} (\vec{k}_1, \vec{k}_2), 
\numberthis\label{eq:f2tilde} \\ 
&\widetilde{G}_2(\vec{k}_1, \vec{k}_2)  = \,\frac{1}{11} \alpha (\vec{k}_1, \vec{k}_2) \left( \widetilde{G}_1 (\vec{k}_1) + \widetilde{F}_1(\vec{k}_2)\right) + \frac{8}{33} \beta (\vec{k}_1, \vec{k}_2) \left(\widetilde{G}_1 (\vec{k}_1) + \widetilde{G}_1 (\vec{k}_2)\right)\\
& -\frac{8}{33} \,c_s^\delta\, k^2 F_2(\vec{k}_1, \vec{k}_2) +\frac{8}{33} \,c_s^\theta\, k^2 G_2(\vec{k}_1, \vec{k}_2) 
+\frac{8}{33} (c_s^\delta - c_s^\theta) \left(\vec{k}\cdot\vec{k}_2\right)  - \frac{8}{33} \sum_{n=1}^4 (c_n^{\delta\delta} - c_{n}^{\delta\theta}-c_n^{\theta\delta}+c_n^{\theta\theta})\, k_ik_j\,e_n^{ij}(\vec{k}_1, \vec{k}_2),  
\numberthis\label{eq:g2tilde} \end{align*}
where $\vec{k}=\vec{k}_1+\vec{k}_2$. Terms in the first lines of $\widetilde{F}_2$ and $\widetilde{G}_2$ come from propagating lower order counterterms. Finally, at third order we find
 \begin{figure*}[t]
\includegraphics[width=0.55 \textwidth]{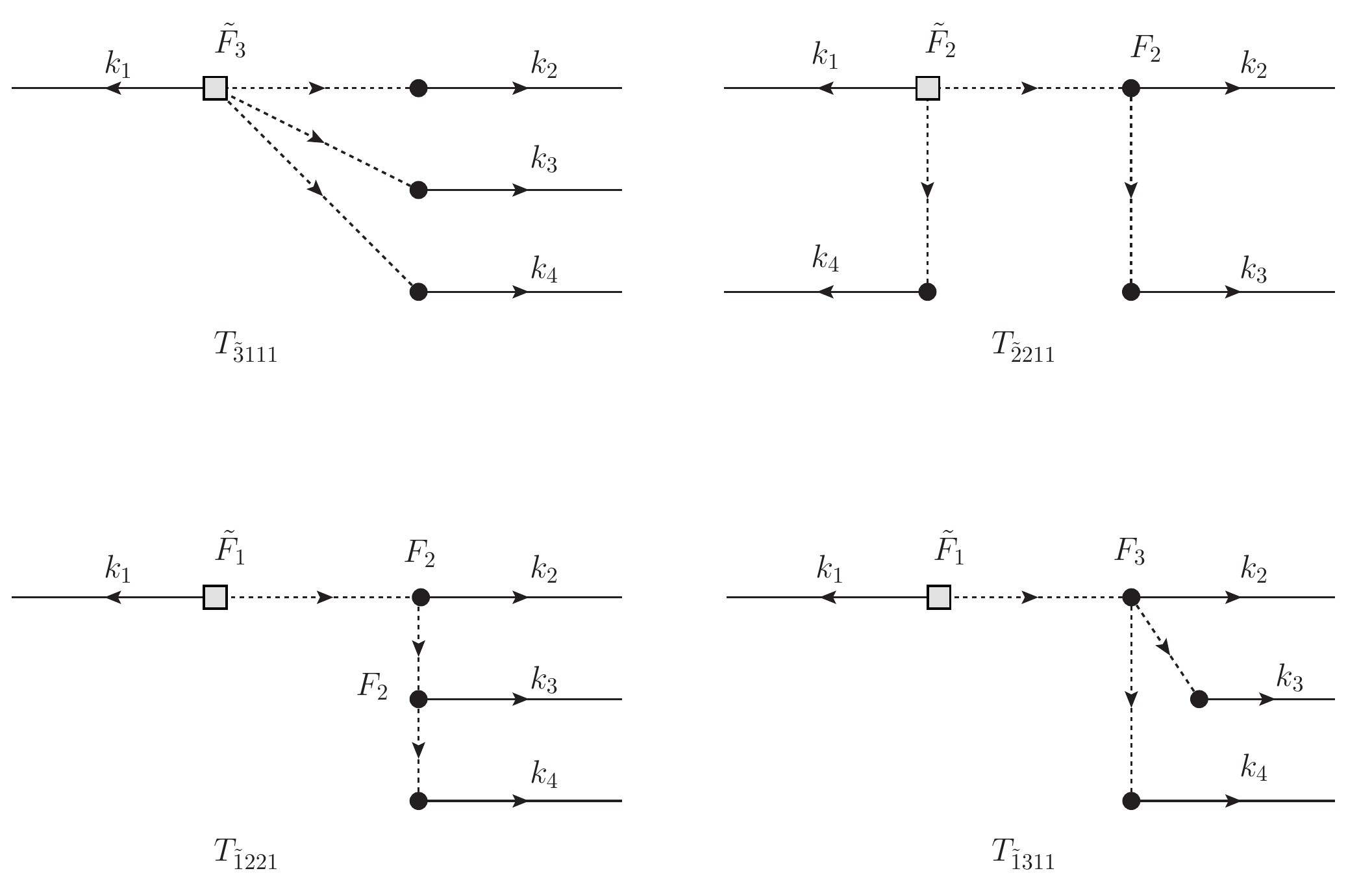}
\caption{Leading EFT contributions to the trispectrum, represented diagrammatically.}
\label{fig:EFT}
\end{figure*}

\begin{align*}
&\widetilde{F}_3(\vec{k}_1,\vec{k}_2,\vec{k}_3) =  \, \frac{11}{52} \alpha( \vec{k}_1,  \vec{k}_2+ \vec{k}_3) \left[ \widetilde{G}_1 ( \vec{k}_1) F_2( \vec{k}_2,  \vec{k}_3) + \widetilde{F}_2 ( \vec{k}_2,  \vec{k}_3) \right] + \frac{11}{52} \alpha( \vec{k}_1+ \vec{k}_2,  \vec{k}_3) \Big[ \widetilde{G}_2( \vec{k}_1,  \vec{k}_2)\\
&+G_2( \vec{k}_1,  \vec{k}_2) \widetilde{F}_1(\vec{k}_3)\Big] + \frac{1}{26} \beta( \vec{k}_1,  \vec{k}_2 +  \vec{k}_3) \left[\widetilde{G}_1 ( \vec{k}_1) G_2( \vec{k}_2,  \vec{k}_3) + \widetilde{G}_2 ( \vec{k}_2,  \vec{k}_3) \right]\\
& + \frac{1}{26} \beta( \vec{k}_1+  \vec{k}_2 ,  \vec{k}_3) \left[ \widetilde{G}_2( \vec{k}_1,  \vec{k}_2) +G_2( \vec{k}_1,  \vec{k}_2) \widetilde{G}_1(\vec{k}_3)\right]+\frac{1}{26} \beta^i_\omega (\vec{k}_1+ \vec{k}_2 , \vec{k}_3)\, \widetilde{G}_{2i}^{\omega}(\vec{k}_1, \vec{k}_2)-\frac{11}{52} \alpha_\omega^i (\vec{k}_1 + \vec{k}_2, \vec{k}_3)\, \widetilde{G}_{2i}^{\omega} (\vec{k}_1, \vec{k}_2)\\
& - \frac{1}{26 } c_s^\delta k^2 F_3(\vec{k}_1, \vec{k}_2, \vec{k}_3) +\frac{1}{26 } c_s^\theta k^2 G_3(\vec{k}_1, \vec{k}_2, \vec{k}_3)+\frac{1}{26} (c_s^\delta F_2(\vec{k}_2, \vec{k}_3)  -c_s^\theta G_2(\vec{k}_2, \vec{k}_3)  ) (\vec{k} \cdot (\vec{k}_2+\vec{k}_3))\\
&-\frac{1}{26} (c_s^\delta-c_s^\theta) (1 - F_2(\vec{k}_1, \vec{k}_2) ) (\vec{k}\cdot\vec{k}_3)
+\frac{1}{26} \sum_{n=1}^4 (c_n^{\delta\delta} - c_{n}^{\delta\theta}-c_n^{\theta\delta}+c_n^{\theta\theta}) \,k_i(k_2+k_3)_j\,e_n^{ij} (\vec{k}_2, \vec{k}_3)\\ 
&-\frac{1}{26}  \sum_{n=1}^4 \Big(c^{\delta\delta}_n  F_2(\vec{k}_2, \vec{k}_3) +(c_n^{\theta\theta} - c_n^{\delta\theta}-c_n^{\theta\delta})G_2( \vec{k}_2, \vec{k}_3)\Big)  k_ik_j\,e_n^{ij}(\vec{k}_1, \vec{k}_2+ \vec{k}_3)\\
&- \frac{1}{26}  \sum_{n=1}^4\Big((c_n^{\delta\delta}-c_n^{\delta\theta}-c_n^{\theta\delta}) F_2(\vec{k}_1, \vec{k}_2)
+c_n^{\theta\theta} G_2(\vec{k}_1, \vec{k}_2) \Big)\,k_ik_j\,e_n^{ij} (\vec{k}_1+ \vec{k}_2, \vec{k}_3)-\frac{1}{26}  \sum_{n=1}^{10} c^{\delta\delta\delta}_n \,k_ik_j E_n^{ij} (\vec{k}_1, \vec{k}_2, \vec{k}_3),\numberthis \label{eq:f3tilde}
\end{align*}
where again $\vec{k}=\vec{k}_1+\vec{k}_2+\vec{k}_3$. The kernel $\widetilde{G}_{2i}^{\omega}$ gives the EFT second order contribution to the vorticity (see \Eq{eq:kernels}), and it is given by
\begin{align*}
&\widetilde{G}_{2i}^{\omega}(\vec{k}_1,\vec{k}_2)= -\frac{2}{9} \epsilon_{ijm} k^j\sum_{n=1}^4 (c_n^{\delta\delta} - c_n^{\delta\theta} - c_n^{\theta\delta} +c_n^{\theta\theta})\,k_l\,e_n^{lm}(\vec{k}_1, \vec{k}_2). \numberthis\label{eq:Gomega}
\end{align*}
For completeness, both in \Eqs{eq:f2tilde}{eq:g2tilde} and \Eq{eq:f3tilde} we have included the expansion of the $(1 + \delta)^{-1}$ term which appears in the equations of motion. Note that the kernels in Eqs.~(\ref{eq:f2tilde}-\ref{eq:Gomega}) are not symmtric in their arguments.

Finally, the amplitudes corresponding to the four EFT trispectrum diagrams shown in Fig.~\ref{fig:EFT}, in terms of symmetrized kernels, are given by
\begin{align*} 
T_{{\tilde 3} 111} &=3! \int \dbar^{\,3}q \, {\widetilde F}_3^s(\vec{k}_2,\vec{k}_3,\vec{k}_4) P_L(q) P_L(k_2) P_L(k_3) P_L(k_4)+3\text{ perms}   \,, \numberthis\\
T_{{\tilde 2
} 211}& = (2!)^2  \int \dbar^{\,3}q \, {\widetilde  F}_2^s(\vec{k}_2+\vec{k}_3, \vec{k}_4) F_2^s(-\vec{k}_2-\vec{k}_3, \vec{k}_3) P_L (q) P_L(\abs{\vec{k}_2 + \vec{k}_3}) P_L(k_3) P_L(k_4) +23\text{ perms}  \,, \numberthis \\
 T_{{\tilde 1} 221} &= (2!)^2  \int \dbar^{\,3}q \, {\widetilde F}_1( -\vec{k}_1) F_2^s( \vec{k}_1 ,  \vec{k}_3+ \vec{k}_4) F_2^s(-\vec{k}_3-\vec{k}_4, \vec{k}_4) P_L(q) P_L(k_1) P_L(\abs{\vec{k}_3+\vec{k}_4}) P_L(k_4)+23\text{ perms} \,,  \numberthis\\ 
 T_{{\tilde 1} 311} &= 3! \int \dbar^{\,3}q \, {\widetilde F}_1(  -\vec{k}_1) F_3^s( \vec{k}_1,  \vec{k}_3,  \vec{k}_4 ) P_L(q) P_L(k_1) P_L(k_3)P_L(k_4) +11\text{ perms} \,. \numberthis \end{align*}
 
As for the case of the SPT amplitudes, we have included a symmetry factor to account for the degenerate configurations of the diagram, and the number of inequivalent permutations of external labels one has to sum over.

\section{\texttt{FnFast}: A Framework to Compute Diagrams in SPT and Beyond}
\label{appFnFast}

The calculation of the one-loop covariance requires the nontrivial evaluation of several diagrams.  Although complex, the computational framework of all of these diagrams is essentially universal, and common to other calculations such as the power spectrum, bispectrum, and the full trispectrum.  It even extends beyond SPT to extensions such as the EFT of LSS, LPT, and RegPT.  We find a need for a computational tool to efficiently represent and evaluate perturbative calculations across a range of theories, one that can be useful not only to those performing these calculations but to those using them in comparisons with simulations and data.  \texttt{FnFast} is a step in this direction.

Perturbative calculations in SPT and similar theories can be represented by diagrams, or graphs.  The evaluation of these diagrams follow some basic rules:
\begin{itemize}
\item Propagators (or edges) receive weights that are scalar functions of the magnitude of momentum flowing in the line (note that here we will use the words momentum and wavenumber interchangeably).
\item Vertices receive weights that are scalar functions of all of the momenta flowing in/out of the vertex.
\item The internal lines are all equivalent and produce no secondary vertices (calculations are performed for a single vertex multiplicity, i.e. $k$-point and $n$-point diagrams do not mix).
\item Symmetry factors, used to count degenerate diagrams, depend only on the topology of the graph.
\item Loop momenta may be efficiently integrated over using Monte Carlo importance sampling methods such as \texttt{VEGAS}.
\end{itemize}
The goal of {\tt FnFast} is to enable a fast, flexible framework to implement these calculations, such that the work required in adding new diagrams, new models, or making use of existing ones, is minimal.  The design philosophy of {\tt FnFast} is that the computationally complex parts of calculations, the evaluation of diagrams, are exposed to the user via simple interfaces.  Users have the freedom to use these diagrams to create their own analyses.  For example, a user may use {\tt FnFast} to evaluate the bispectrum diagrams for a range of cosmologies, and then use the results to perform fits with a separate analysis code.

{\tt FnFast} is written in {\tt C++11} and makes use of the external libraries {\tt gsl} (for interpolation) and {\tt CUBA} (for integration).  The code is publicly available at {\color{blue}\href{https://github.com/jrwalsh1/FnFast}{{\tt github}}}.

The name {\tt FnFast} derives from the need to have fast evaluation of the SPT kernels $F_n$ and $G_n$ when performing the calculation of the one-loop covariance.  This calculation calls $F_5$, which is a very long function of the momenta in the diagram.  Recursive evaluation of this function can be extremely inefficient, and repeated evaluations can be avoided through dynamic programming, precomputation of combinatoric ingredients, and opportunistic caching.  To speed up the kernel computation, we have found (probably not for the first time) a form of the SPT recursion relations that are very amenable to efficient numeric evaluation, which we derive in Appendix~\ref{recursionrelations}.

\subsection{Overview of the Code}
\label{codestructure}

\begin{figure*}[!htb]
\begin{center}
\includegraphics[width=0.7\textwidth]{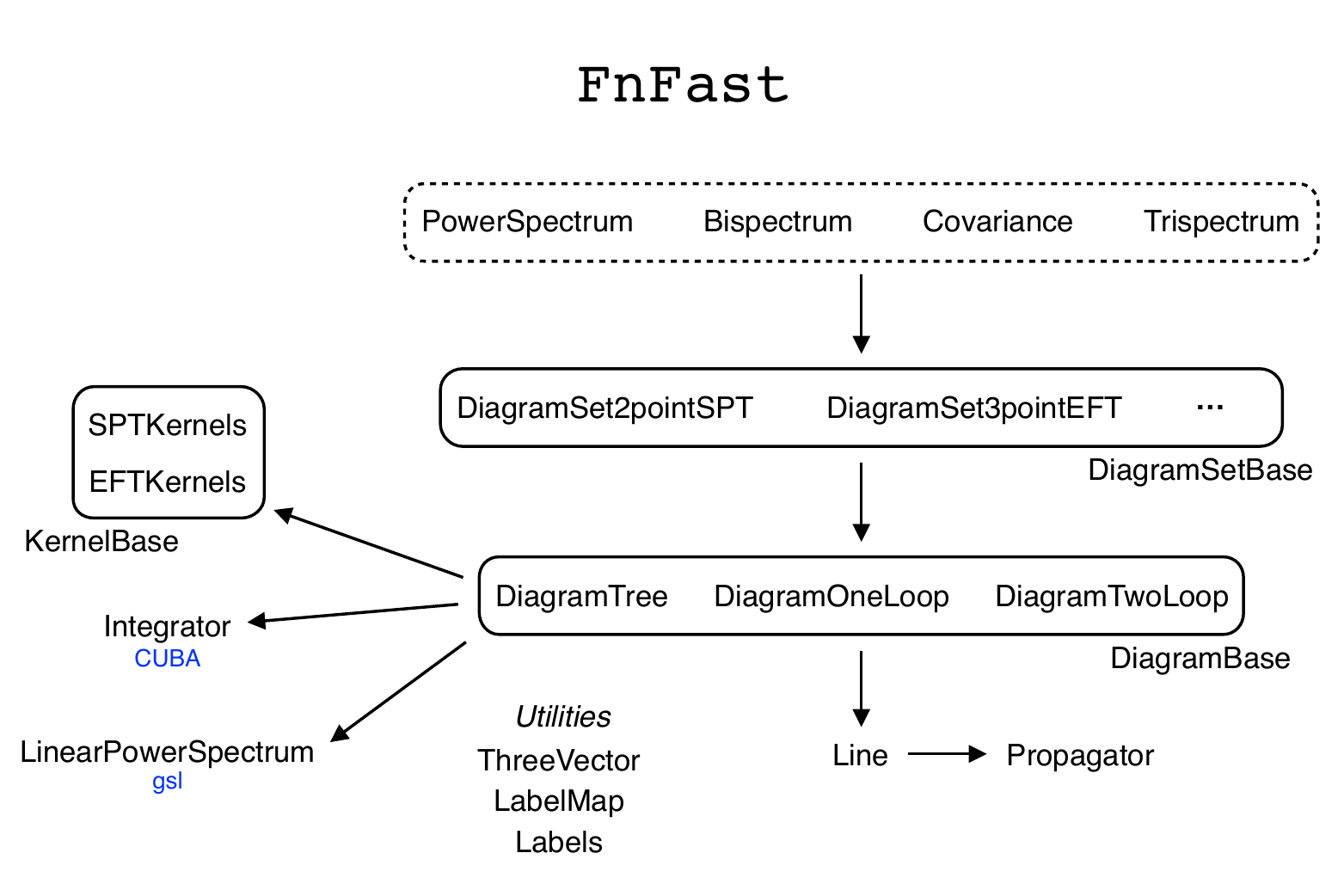}
\caption{Organization of the \texttt{FnFast} code.}
\label{FnFast}
\end{center}
\end{figure*}

We briefly describe the structure of {\tt FnFast} and how calculations are performed.  Figure~\ref{FnFast} depicts the salient structure of the code.  
The code is organized so that diagrams are constructed independent of any linear power spectrum or momentum.  Diagrams are encoded via their functional dependence on these objects, freeing the user from specifying any cosmological parameters or explicit momenta until evaluation of the diagrams takes place.  Additionally, this avoids code duplication and avoids typical users from having to interact with the code performing the explicit evaluation of diagrams.  This is primarily accomplished via a labeling system that maps labels representing vertices and momenta to generic objects (e.g., kernels or linear power spectra).  The {\tt LabelMap} object functions as an associative array; for example:
\begin{lstlisting}
   LabelMap<A, B> label_map {{a1, b1}, {a2, b2}};
\end{lstlisting}
maps objects {\tt a1} and {\tt a2} of type {\tt A} to {\tt b1} and {\tt b2} of type {\tt B}, respectively.  This allows one to simply construct diagrams in terms of linear power spectrum and kernel objects without explicit instances of these objects.

Tree and one-loop diagrams (and also two-loop diagrams in future releases) are defined as instances of (distinct) classes that know how to evaluate them, including symmetry factors, momentum permutations, and IR regulation.  Each diagram is evaluated only at the fully exclusive level, with the complete momentum dependence specified; higher levels of the calculation are responsible for integration over momentum components.

All diagram objects are derived from a common {\tt DiagramBase} object.  Sets of diagrams are grouped into a {\tt DiagramSetBase} object, which allows one to collect diagrams together for efficient evaluation.  Currently, the code contains the following sets of diagrams:
\begin{itemize}
\item 2-, 3-, and 4-point diagrams in SPT at tree and one-loop level.
\item 2-, 3-, and 4-point diagrams in the EFTofLSS at one-loop level.
\end{itemize}
Calculations use these diagram sets and perform necessary phase space integrals or other analysis.  This set of diagrams is sufficient to do many calculations, and the framework is set up to make it simple to add new models or diagrams.

\subsection{Constructing Diagrams}
\label{FnFast:construct}

{\tt FnFast} represents diagrams by their topology, independent of any assignment of vertex rules, momenta, or linear power spectra.  This makes it straightforward to build all of the diagrams for a calculation, and the functional dependence on momentum and linear power spectrum is injected automatically at evaluation time, making it easy to vary cosmological parameters and obtain predictions.  

As an example, here is code to construct the $T_{4211b}$ diagram:
\begin{lstlisting}
      // T4211b
      // vertex types
      LabelMap<Vertex, VertexType> vertex_types 
         {{Vertex::v1, VertexType::type1}, {Vertex::v2, VertexType::type1},
          {Vertex::v3, VertexType::type1}, {Vertex::v4, VertexType::type1}};
      // kernel types
      LabelMap<Vertex, KernelType> kernel_types 
         {{Vertex::v1, KernelType::delta}, {Vertex::v2, KernelType::delta}, 
          {Vertex::v4, KernelType::delta}, {Vertex::v4, KernelType::delta}};
      // momentum flow
      LabelMap<Momentum, Propagator::LabelFlow> mom_q {{Momentum::q, Propagator::LabelFlow::Plus}};
      LabelMap<Momentum, Propagator::LabelFlow> mom_qk2 {{Momentum::q, Propagator::LabelFlow::Minus}, {Momentum::k2, Propagator::LabelFlow::Plus}};
      LabelMap<Momentum, Propagator::LabelFlow> mom_k3 {{Momentum::k3, Propagator::LabelFlow::Plus}};
      LabelMap<Momentum, Propagator::LabelFlow> mom_k4 {{Momentum::k4, Propagator::LabelFlow::Plus}};
      // propagators
      Propagator prop_T4211b_q(mom_q);
      Propagator prop_T4211b_qk2(mom_qk2);
      Propagator prop_T4211b_k3(mom_k3);
      Propagator prop_T4211b_k4(mom_k4);
      // lines
      Line line_T4211b_12a(Vertex::v1, Vertex::v2, prop_T4211b_q);
      Line line_T4211b_12b(Vertex::v1, Vertex::v2, prop_T4211b_qk2);
      Line line_T4211b_13(Vertex::v1, Vertex::v3, prop_T4211b_k3);
      Line line_T4211b_14(Vertex::v1, Vertex::v4, prop_T4211b_k4);
      std::vector<Line> lines_T4211b {line_T4211b_12a, line_T4211b_12b, line_T4211b_13, line_T4211b_14};
      // define the diagram
      DiagramOneLoop T4211b(lines_T4211b, vertex_types, kernel_types);
\end{lstlisting}

The {\tt Propagator} object represents algebraic dependence on a set of momenta, specified via the {\tt Momentum} label its direction (the {\tt LabelFlow} object).  A {\tt Line} associates the beginning and ending vertices in a graph with the propagator, and the diagram is built from lines.  Additionally, one can specify whether the $N$-point function is built from $\delta$ or $\theta$ correlators, and one also uses the {\tt VertexType} label to specify whether vertices in the graph should be considered distinct or identical. We note that the {\tt VertexType} label values are purely dummy -- they are only used to denote whether different vertices in the same graph are equivalent or not, which is necessary in calculating the symmetry factor and momentum permutations for a diagram.

\subsection{Evaluating Diagrams}
\label{FnFast:evaluate}

Currently, tree-level and one-loop calculations for 2-, 3-, and 4-point diagrams can be performed, and support for two-loop calculations is under development.  At one loop, the IR regulation procedure described in Ref.~\cite{2014JCAP...07..056C} is provided to automatically render the integrand IR finite.  Kernels are provided for SPT and the EFTofLSS, with support for additional models planned.  

Since calculations evaluate sets of diagrams, it is convenient to bundle the necessary diagrams for a calculation into a {\tt DiagramSet} object.  The base class defines a thin wrapper that will evaluate a set of fully exclusive diagrams and sum the results; derived instances can also provide functions such as the value of individual diagrams.  We find that it is ideal to put the actual diagram definitions into {\tt DiagramSet} classes, which gives the diagrams a simple interface for evaluation and separates them from the code that defines how they are evaluated.  For example, the above code snippet for $T_{4211b}$ lives in the {\tt DiagramSet4PointSPT} class.

Integration over loop momenta or unobserved components of external momenta for diagrams is defined at a higher level, namely in calculations.  For example, the {\tt Covariance} calculation defines how the various 4-point diagrams are used to compute the covariance, including how phase space is sampled, what the limits of integration are, and what levels of the calculation are exposed to the user.  These are the calculations that those wishing to work with existing calculations will be using or defining; the code is structured so that building these calculations does not require the user to interact with the implementation of the diagrams.  The integration routines in {\tt CUBA} are also largely separated from the user; building a calculation simply requires
\begin{itemize}
\item Defining how phase space is sampled.
\item Defining the integrand.
\item Defining the limits of integration.
\end{itemize}
For example, the main function performing the calculation one-loop SPT contribution to the covariance is
\begin{lstlisting}
   // integration method
   PhaseSpace phasespace(k, kprime, _UVcutoff, &kernels, PL, this);
   phasespace.ndim = 4;

   // VEGAS integration via cuba
   VEGASintegrator vegas(phasespace.ndim);

   return vegas.integrate(oneLoop_integrand, &phasespace);
\end{lstlisting}

The phase space is constructed with explicit $k$ and $k'$ magnitudes for momenta in the covariance, the UV cutoff for the integrals, the object providing the SPT kernels, the object providing the linear power spectrum, and the {\tt this} pointer that passes the instance of the {\tt Covariance} calculation into the phase space (the {\tt PhaseSpace} object uses it to evaluate the relevant diagrams, which the {\tt Covariance} controls)\footnote{The integration library {\tt CUBA}, which is both powerful and versatile, uses a {\tt C}-style interface to define integrands that is restrictive for object-oriented design.  The use of the {\tt this} pointer was one of the more parsimonious approaches to define integrands that we found.}.  The integration routines in VEGAS are then simply called and will evaluate the integrand to a prescribed/default accuracy.

\subsection{An Efficient Form of the SPT Recursion Relations}
\label{recursionrelations}

The recursion relations for the SPT kernels $F_n$ and $G_n$ may be written
\begin{align} \label{eq:FGrecursion}
F_n (\bq_{1..n}) &= \sum_{k=1}^{n-1} G_k (\bq_{1..k}) \Bigl[ c_{F,\alpha}^{(n)} \alpha(\bq^{\Sigma}_{1..k}, \bq^{\Sigma}_{k+1..n}) F_{n-k} (\bq_{k+1..n}) + c_{F,\beta}^{(n)}  \beta(\bq^{\Sigma}_{1..k}, \bq^{\Sigma}_{k+1..n}) G_{n-k} (\bq_{k+1..n})\Bigr] \,, \nn \\
G_n (\bq_{1..n}) &= \sum_{k=1}^{n-1} G_k (\bq_{1..k}) \Bigl[ c_{G,\alpha}^{(n)} \alpha(\bq^{\Sigma}_{1..k}, \bq^{\Sigma}_{k+1..n}) F_{n-k} (\bq_{k+1..n}) + c_{G,\beta}^{(n)} \beta(\bq^{\Sigma}_{1..k}, \bq^{\Sigma}_{k+1..n}) G_{n-k} (\bq_{k+1..n}) \Bigr] \,,
\end{align}
where $\bq_{1..k} = \bq_1, \bq_2, \ldots, \bq_k$ in the arguments of $F$ and $G$, and $\bq^{\Sigma}_{1..k} = \sum \bq_{1..k}$ is the sum of these momenta in the arguments of $\alpha$ and $\beta$.  The coefficients $c_{\{F, G\}, \{\alpha, \beta\}}^{(n)}$ are simple rational functions of $n$.  This recursion relation, while simple, is inefficient for evaluation since different branches of the recursion for the symmetrized kernels (whose form we are ultimately interested in) will frequently reevaluate the same $\alpha$ and $\beta$ kernels.  To avoid this, one can write the symmetrized kernels grouped by the $\alpha$ and $\beta$ functions, symmetrizing over the arguments of $F$ and $G$ in the recursion relation.

First, we write the compact notation
\begin{align}
\text{first $k$ momenta in } \pi(\bq_{1..n})&: \; \bq^{\pi, L}_{k, n} \nonumber \\
\text{last $n-k$ momenta in } \pi(\bq_{1..n})&: \; \bq^{\pi, R}_{k, n}
\end{align}
where $\pi(\bq_{1..n})$ is some permutation of the momenta in $\bq_{1..n}$, and we will assume the sum is implied when using this form in the $\alpha$ and $\beta$ kernel arguments.  The recursion relation for $F_n$ becomes
\begin{align}
F_n (\bq_{1..n}) &= \sum_{k=1}^{n-1} G_k (\bq^{{\bf 1}, L}_{k, n}) \Bigl[ c_{F,\alpha}^{(n)} \alpha(\bq^{{\bf 1}, L}_{k, n}, \bq^{{\bf 1}, R}_{k, n}) F_{n-k} (\bq^{{\bf 1}, R}_{k, n}) + c_{F,\beta}^{(n)} \beta(\bq^{{\bf 1}, L}_{k, n}, \bq^{{\bf 1}, R}_{k, n}) G_{n-k} (\bq^{{\bf 1}, R}_{k, n}) \Bigr] \,,
\end{align}
where $\bf 1$ is the identity, and similarly for $G_n$.  The symmetrized kernel is therefore
\begin{align}
F_n^{s} (\bq_{1..n}) &= \frac{1}{n!} \sum_{\pi \in \sigma_n} \sum_{k=1}^{n-1} G_k (\bq^{\pi, L}_{k, n}) \Bigl[ c_{F,\alpha}^{(n)} \alpha(\bq^{\pi, L}_{k, n}, \bq^{\pi, R}_{k, n}) F_{n-k} (\bq^{\pi, R}_{k, n}) + \beta(\bq^{\pi, L}_{k, n}, \bq^{\pi, R}_{k, n}) c_{F,\beta}^{(n)} G_{n-k} (\bq^{\pi, R}_{k, n}) \Bigr] \,,
\end{align}
There are $n!$ permutations of the $n$ momenta, but for a given $k$ there are only $\binom{n}{k}$ distinct divisions of these momenta into specific groups of $k$ and $n-k$ momenta.  Therefore, there are equivalence classes of permutations that will combine to symmetrize the kernels inside the sum; the result is the recursion relation
\begin{align}
F_n^{s} (\bq_{1..n}) &=  \sum_{k=1}^{n-1} \binom{n}{k}^{-1} \sum_{\pi \in \sigma_n^{k, {\rm ord}}} G^{s}_k (\bq^{\pi, L}_{k, n}) \Bigl[ c_{F,\alpha}^{(n)} \alpha(\bq^{\pi, L}_{k, n}, \bq^{\pi, R}_{k, n}) F^{s}_{n-k} (\bq^{\pi, R}_{k, n}) + \beta(\bq^{\pi, L}_{k, n}, \bq^{\pi, R}_{k, n}) c_{F,\beta}^{(n)} G^{s}_{n-k} (\bq^{\pi, R}_{k, n}) \Bigr] \,.
\end{align}
The set of permutations $\sigma_n^{k, {\rm ord}}$ are those where the two subsets of $k$ and $n-k$ indices are each ordered.  For example, for $k = 1$ and $n = 4$, the permutation $\{3, 1, 2, 4\}$ is included (which divides into the ordered subsets $\{3\}$ and $\{1, 2, 4\}$), but $\{2, 1, 4, 3\}$ is not.

This is a recursion relation directly on the symmetrized kernels.  Using the base cases $F_0 = G_0 = 0$ and $F_1 = G_1 = 1$, and using the symmetry between $\sigma_n^{k, {\rm ord}}$ and $\sigma_n^{n-k, {\rm ord}}$, for a given $n$ there are only $2^{n-1} - 1$ distinct momentum combinations at first step in the recursion for which we need to compute $F^{s}$ and $G^{s}$ (instead of the na\"ive $n!$).  Tracing through the entire recursion relation to the base cases, $F_n^{s}$ requires only $(n-1)!!$ distinct evaluations of symmetrized kernels.  The na\"ive implementation would evaluate $2^{n-1} (n-1)!!$ kernels, meaning this approach can be significantly faster than the na\"ive one.  Additionally, since $n \leq 7$ for any practical application, we can (and in {\tt FnFast} do) pre-compute any combinatoric objects, such as the permutations summed over, that will be repeatedly used in the recursion relations.

\twocolumngrid

\bibliographystyle{revtex}

\bibliography{cov_short}
\end{document}